\documentclass{jpp}
\usepackage{graphicx}

\usepackage[utf8]{inputenc}
\usepackage[T1]{fontenc}
\usepackage{amsmath}
\usepackage{xcolor}
\usepackage{hyperref}
\usepackage{subfig}
\usepackage{xspace}
\usepackage{titlesec}
\usepackage{layouts}
\usepackage[nolist,nohyperlinks]{acronym}

\titleformat{\paragraph}
{\normalfont\normalsize}{\theparagraph}{1em}{}
\titlespacing*{\paragraph}
{0pt}{3.25ex plus 1ex minus .2ex}{1.5ex plus .2ex}

\def\nsbh{\ac{NS}-\ac{BH}\xspace}
\def\bns{\ac{NS}-\ac{NS}\xspace}
\def\bbh{\ac{BH}-\ac{BH}\xspace}

\shorttitle{EM counterparts of CBM}
\shortauthor{S. Ascenzi, G. Oganesyan, M. Branchesi, R. Ciolfi}

\title{Electromagnetic Counterparts of Compact Binary Mergers}

\author{Stefano Ascenzi\aff{1,2,3}
  \corresp{\email{ascenzi@ice.csic.es}},
  Gor Oganesyan\aff{4,5},\\
  Marica Branchesi\aff{4,5}, 
  Riccardo Ciolfi\aff{6,7}}

\affiliation{\aff{1}INAF -- Osservatorio Astronomico di Brera, Via E. Bianchi 46, I-23807 Merate, Italy
\aff{2} Institute of Space Sciences (ICE, CSIC), Campus UAB, Carrer de Can Magrans s/n, 08193, Barcelona, Spain
\aff{3} Institut d’Estudis Espacials de Catalunya (IEEC), Carrer Gran Capita 2–4, 08034 Barcelona, Spain
\aff{4} Gran Sasso Science Institute, Viale F. Crispi 7, I-67100, L’Aquila (AQ), Italy
\aff{5} INFN - Laboratori Nazionali del Gran Sasso, I-67100, L’Aquila (AQ), Italy
\aff{6} INAF–Osservatorio Astronomico di Padova, Vicolo dell’Osservatorio 5, I-35122 Padova, Italy
\aff{7} INFN–Sezione di Padova, Via Francesco Marzolo 8, I-35131 Padova, Italy}

\begin{document}

\maketitle

\begin{abstract}
The first detection of a binary neutron star merger through gravitational waves and photons marked the dawn of multi-messenger astronomy with gravitational waves, and it greatly increased our insight in different fields of astrophysics and fundamental physics. However, many open questions on the physical process involved in a compact binary merger still remain and many of these processes concern plasma physics. 
With the second generation of gravitational wave interferometers approaching their design sensitivity, the new generation under design study, and new X-ray detectors under development,
the high energy Universe will become more and more a unique laboratory for our understanding of plasma in extreme conditions. In this review, we discuss the main electromagnetic signals expected to follow the merger of two compact objects highlighting the main physical processes involved and some of the most important open problems in the field.
\end{abstract}

\begin{acronym}
\acrodef{BH}[BH]{black hole}
\acrodef{EM}[EM]{electromagnetic}
\acrodef{EOS}[EOS]{equation of state}
\acrodef{GRB}[GRB]{gamma-ray burst}
\acrodef{GW}[GW]{Gravitational wave}
\acrodef{ISCO}[ISCO]{innermost stable circular orbit}
\acrodef{KAGRA}[KAGRA]{Kamioka Gravitational wave detector}
\acrodef{NS}[NS]{neutron star}
\acrodef{SGRB}[SGRB]{Short \ac{GRB}}
\acrodef{LGRB}[LGRB]{Long \ac{GRB}}
\acrodef{ShortGRB}[SGRB]{Short \ac{GRB}}
\acrodef{SNR}[S/N]{signal-to-noise ratio}
\acrodef{KN}[KN]{kilonova}
\acrodefplural{KN}[KNe]{kilonova}
\acrodef{CBM}[CBM]{compact binary merger}
\acrodef{MHD}[MHD]{magneto-hydrodynamical}
\acrodef{SMNS}[SMNS]{supramassive neutron star}
\acrodef{HMNS}[HMNS]{hypermassive neutron star}
\acrodef{SDPT}[SDPT]{spindown-powered transient}
\acrodef{PWN}[PWN]{pulsar wind nebula}
\acrodefplural{PWN}[PWNe]{pulsar wind nebula}
\acrodef{MRI}[MRI]{magneto-rotational instability}
\acrodef{BZ}[BZ]{Blandford–Znajek}
\acrodef{BATSE}[BATSE]{Burst \& Transient Source Experiment}
\acrodef{CGRO}[CGRO]{Compton Gamma Ray Observatory}
\acrodef{LVC}[LVC]{Ligo and Virgo Scientific Collaborations}
\acrodef{NIR}[NIR]{Near-Infrared}
\end{acronym}

\section{Introduction}

\acp{GW} are perturbations of the spacetime propagating as waves at the speed of light. Although their interaction with matter is weak, extreme astrophysical processes are able to release through \acp{GW} such a huge amount of energy to be detectable by us. A class of such extreme astrophysical events is the merger of two compact objects -- namely objects with a huge mass compressed in a very small volume -- orbiting one around the other, where the emission of \acp{GW} removes angular momentum from the system shrinking the orbital separation between the two objects and leading to their ultimate coalescence. Examples of compact objects are \acp{NS} and \acp{BH}, for which a mass of the order of a Solar mass is compressed within a radius of $\sim\!10$\,km or less. It was indeed the merger of a system formed by two \acp{BH} the first \ac{GW} event ever detected, which has been observed on 14th September 2015 by the \ac{LVC} using the LIGO laser interferometers \citep{GW150914}. The discovery was followed in the subsequent years by other detections of \bbh mergers \citep{BBHO1, GW151226, GW170104, GW170608, GW170814, NewBBH}.
The 17th August 2017 a different signal, originated by the coalescence of two \acp{NS} has been observed by the LIGO and Virgo interferometer network \citep{GW170817}. This detection was particularly interesting because, while for a \bbh merger the emission of photons is not generally expected, the coalescence of a \bns system should be accompanied by a copious amount of \ac{EM} radiation. Indeed, few seconds after the merger, the satellites FERMI and INTEGRAL detected a flare of $\gamma$-rays in a direction of the sky compatible with the arrival direction of the \acp{GW} \citep{Goldstein2017, Savchenko2017}. A few hours later, ground-based telescopes identified the host galaxy of the \ac{GW} source, NGC4993, by detecting the optical and \ac{NIR} counterpart (e.g.,  \citealt{Coulter2017, PianDavanzo2017, DroutPiro2017, Chornock2017, Shappee2017}). Few days later the source was observed also in the X-ray and radio bands, and the observations continued for years after the \ac{GW} event \citep{TrojaPiro2017, Haggard2017, AlexanderMargutti2018, DavanzoCampana2018, MooleyDeller2018, Ghirlanda2019, Hajela2020, Troja2020}. 
This \bns merger, known as GW170817, which was observed through gravitational and multi-wavelength \ac{EM} radiation \citep{GW170817MMA}, marked the dawn of the new field of multimessenger astronomy with \acp{GW}. 

The impact of the first \ac{GW}-\ac{EM} multimessenger detection has been wide. It contributed to answer many fundamental questions within the field of high-energy astrophysics and fundamental physics, such as the origin of a class of very energetic $\gamma$-ray emissions coming from distant galaxies known as \emph{short $\gamma$-ray bursts}, and the existence of optical sources powered by the radioactive decay of heavy elements, known as \emph{kilonovae} (or macronovae).  
\bns mergers reveal to be able to form collimated and very energetic outflows of matter, and to be important (likely dominant) formation sites of the heaviest elements in the Universe such as gold, platinum, the lanthanides and the actinides. 
Moreover, this event furnished constrains, independent from those already existent, to the unknown equation of state of \ac{NS} \citep{LVC2018_tidal} and the the Hubble constant \citep{Fishbach2019}, extending the range of influence of multimessanger astronomy to the fields of quantum chromodynamics and cosmology. It is evident thus that the scientific potential of multimessenger astronomy is huge and covers many branches of physics. 

\ac{GW} astrophysics and the physics governing the electromagnetic emission from \ac{GW} sources are still in their early phase, and there is still an enormous theoretical effort to be done to be ready to interpret the multi-messenger observations 
expected in the upcoming years. In this context, the physics of plasma plays a key role.
The majority of the problems, which are not fully understood yet, require a deep knowledge of the dynamics of the plasmas in extreme regimes.
Examples include the amplification of magnetic field during a \bns merger, the acceleration of particles responsible for the observed radiation and the role of magnetic reconnection in high-energy astrophysical processes. On the other hand, the Universe constitutes a unique laboratory to get a better insight on the physics of plasma. 
It allows us to observe and study regimes of temperature, density and magnetic field that cannot be reached in laboratory experiments. 
In the near future, we expect great possibilities to explore these regimes due to the recently developed \ac{GW} detectors and their next generation operating in synergy with innovative multi-wavelenght observatories, such as the Cherenkov Telescope Array (CTA) \citep{CTA2011, CTA2013}, the proposed THESEUS mission \citep{Theseus2018,Stratta2018}, the VEra Rubin Observatory \citep{LSST2019}, the European Extremely Large Telescope (E-ELT) \citep{ELT2007}, the Square Kilometre Array (SKA) \citep{SKA2004}, and more. This will boost considerably our understanding of the high-energy Universe.

Here, we review the main \ac{EM} counterparts of compact binary mergers -- in particular \bns and \nsbh mergers -- describing in a synthetic form the leading theoretical models and the most important open problems in the field. Our aim is not to give the reader a complete dissertation of the topic, but to provide a basic insight on the astrophysical processes involved in the coalescence of a compact binary system. For more detailed information, we will refer the interested reader to other reviews devoted to the specific topics summarized here (e.g., \citealt{P05, Shibata2011, B13, Berger2014, KuZh2015,  R15, Baiotti2017, Metzger2017, Ciolfi2018, Nakar2019, Ciolfi2020b}).

This review is organized as follows: In Section \ref{sec:definitions} we clarify some basic definitions useful for the reader. In Section \ref{sec:coalescence} we describe the dynamics of a compact binary merger. In Section \ref{sec:EM} we describe the process occurring after the merger, the \ac{EM} emission they are responsible for, and the open problems in the field. Finally, in Section \ref{sec:final} we present our final remarks.

\section{Basic definitions and constants}
\label{sec:definitions}

In this Section we summarize some basic terms that are commonly used by astrophysics but whom the plasma physics community may not be familiar with. 
\begin{itemize}
    \item Transient:  A transient is an astrophysical phenomenon with a short temporal duration compared to human timescales, namely it can last from seconds (or less) to years. For example, a supernova, which can be observable for months, is a transient, while a non-variable star, whose lifetime lies in the range $10^6-10^{10}$ years, is not.
    \item Metallicity: abundance of elements heavier than Helium (hence with atomic number greater than 2). In astrophysics generally we refer to these elements as \emph{metals}. 
    \item{Compact object: astrophysical object with high mass compressed in a small volume. In particular, here compact object will refer only to \acp{NS} and (stellar-mass) \acp{BH}. A \ac{NS} has a mass equal to 1-2 times the mass of the Sun within a radius of $10-13\,\rm km$. \acp{BH} treated in this review have a mass between few times and 100 times the mass of the Sun. Their Schwarzschild radius increases linearly with the \ac{BH} mass.
    For a \ac{BH} of one solar mass, the Schwarzschild radius is $\sim 3\,\rm km$}.
    \item{Lightcurve: the lightcurve of a transient is the temporal evolution of the luminosity (or flux) of the source in a given frequency range. We refer to bolometric lightcuve when the frequency range coincides with the whole \ac{EM} spectrum.}
\end{itemize}

Throughout the review the masses will be commonly expressed in terms of solar masses, where 
   a solar mass corresponds to $1.989 \times 10^{33}\, \rm g $, and is indicated by the symbol $M_\odot$. Distances of astrophysical sources are expressed in multiples of \emph{parsec} ($\rm pc$), where $1\,\rm pc = 3.09 \times 10^{18}\,\rm cm$.

\section{Coalescence Dynamics}
\label{sec:coalescence}

\begin{figure}
\centering
\includegraphics[width = \columnwidth]{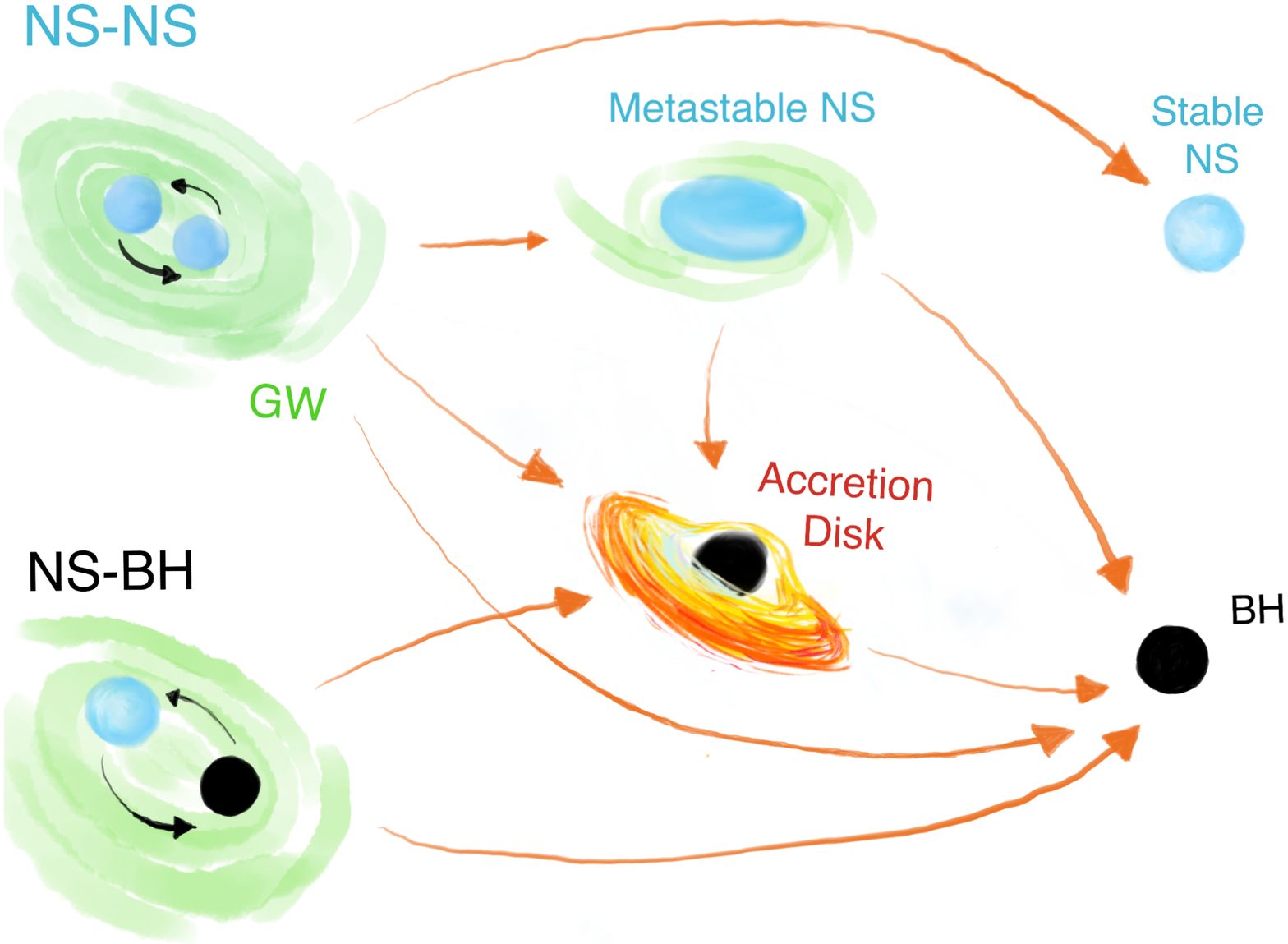}
\caption{Different scenarios for a \bns and a \nsbh merger and the merger remnant. The \ac{EM} radiation is expected when an accretion disk and unbound mass are left outside the merger remnant.
}
\label{fig:decay}
\end{figure}

The merger of a \bns or a \nsbh binary system is a complex event that requires full general relativistic \ac{MHD} simulations to be properly studied. In fact, different initial orbital parameters and/or \ac{NS} \ac{EOS} can lead to a substantially different merger dynamics and in the case of a \bns merger even to different types of remnant objects. All the possible merger channels are summarized in Figure \ref{fig:decay}. 
While a \nsbh merger can only result in the formation of a more massive \ac{BH}, with an accretion disk if the \ac{NS} is disrupted outside the \ac{BH}'s innermost stable circular orbit or without a disk in the opposite case, a \bns merger can result in the formation of a \ac{BH}, a stable \ac{NS} or, in most cases, a metastable \ac{NS} collapsing to a \ac{BH} after some time (e.g., \citealt{ShiTaniUry2005, ShiTan2006, BaGiRe2008, ReBaGi2010, HoKyOk2011, BaBaJa2013, GiaPer2013, CiolfiKastaun2017, Bernuzzi2020}). This last case can be divided in two sub-regimes: if the \ac{NS} can be supported against collapse by uniform rotation the star is called \ac{SMNS} and it remains stable as long as the rotation is not quenched. It is not clear for how long the collapse can be delayed and still result in a \ac{BH} surrounded by a massive accretion disk \citep{MaMeBe2015,Ciolfi2019}. If the star is instead so massive that can only be supported in presence of differential rotation, it will collapse as soon as the NS core has acquired uniform rotation, which happens on timescales of $O(100\, \rm ms)$. This object is named \ac{HMNS}, and it likely leads to the formation of an accretion disk. 

It is worth noting that during a \bns merger general relativistic \ac{MHD} simulations show the development of the Kelvin-Helmholtz (KH) instability in the shear layer separating the two NS cores when they first come into contact. The generation of vortices amplifies the toroidal component of the magnetic field, even by one order of magnitude or more in the first few milliseconds (e.g., \citealt{Kiuchi2015}). Later on, further amplification is provided by magnetic field line winding and the \ac{MRI} \citep{BalbusHawley1991, DuezLiu2006, SiCiHaRe2013}. Combined together, these processes are expected to increase the magnetic field strengths from pre-merger levels of the order of $B\sim 10^{12} \rm G$ up to $B\sim10^{15}-10^{16}\, \mathrm{G}$ (e.g., \citealt{PR06, GiZr2015}). 
A \ac{NS} with such a high magnetic field is usually called \emph{magnetar}
\footnote{In a different context, magnetar refers to a \ac{NS} whose persistent X-ray luminosity and bursting activity are powered by its strong magnetic field (see \citet{Mereghetti2015} for a recent review).  Here, we simply call magnetar a \ac{NS} remnant with a magnetic field $B>10^{14}\,\rm G$. In all the models described here, this huge magnetic field is used to extract the rotational energy of the star, which is the main reservoir of energy that powers the emission.}.
In this context, an important issue is represented by the finite spatial resolution of the simulations, which is currently not sufficient to fully capture small-scale amplification mechanisms like the KH instability \citep{Kiuchi2015}.
To overcome this limitation, effective sub-grid amplification methods have been proposed (\emph{e.g.} \citealt{GiZr2015, Carrasco2020, Vigano2020}).

A further important aspect to explore in \acp{CBM} is the ejection of mass during the merger and post-merger phases. Part of the \ac{NS} matter can expelled and become unbound \citep{Davies1994, RoLi1999}. 
Tidal forces right before merger can cause partial disruption of the (two) NS(s), with material launched at mildly relativistic velocities on the orbital plane of the system (especially for binary systems characterized by objects of unequal mass). Moreover, once the accretion disk is formed, neutrino irradiation, nuclear recombination, and magnetohydrodynamic viscosity can drive mass outflows from the disk (see Sec. \ref{sec:kilonova} for more details and references).
In the specific case of \bns mergers, further ejection of matter in all directions (including polar regions) is produced by the shocks launched in the contact interface between the two stars when they touch. Moreover, the (meta)stable massive \ac{NS} resulting from the merger can launch an additional baryon-loaded wind up to its collapse to a BH, if any (e.g., \citealt{CiolfiKalinani2020}).  
More details on these different ejecta components are given in Sections \ref{sec:kilonova} and \ref{sec:SDPT}. 

We conclude this section by briefly providing a qualitative description of the \ac{GW} signal expected from a \ac{CBM}. This signal can be divided in three phases: the \emph{inspiral}, where the two objects are still distant and can be treated as point-like masses, the \emph{merger}, where the objects come into contact and matter and finite size effects play an important role, and the \emph{ringdown}, the final phase in which the newly formed compact object relaxes in a stationary configuration emitting an exponentially damped oscillating signal. The \ac{GW} waveform of the inspiral part is a signal oscillating at a frequency of $f_{\rm GW} = 2f_{\rm K}$, where $f_{\rm K}$ is the Keplerian orbital frequency of the system which increases in time as the stars get closer. Its amplitude in turn follows the increase of frequency as $f_{\rm GW}^{2/3}$. The resulting waveform is called \emph{chirp} \footnote{It is qualitatively similar to the chirp of a bird.}. 

Contrary to the inspiral signal, detailed simulations in general relativity are required to calculate
the merger and post-merger GW signal, which strongly depend on the (still unknown) \ac{EOS} describing NS matter at supranuclear densities. 
For NS-NS mergers, the post-merger signal (until collapse to a BH, if any) is typically dominated by a single frequency, related to the fundamental oscillation mode of the remnant. 
On longer timescales, if a long-lived \ac{SMNS} or a stable \ac{NS} is formed, it is expected to assume an ellipsoidal shape. This object will then further emit \ac{GW}s at both $2f_{\rm rot}$ and $f_{\rm rot}$ frequencies, where $f_{\rm rot}$ is the spin frequency.
Since the object keeps spinning down and the amplitude of the signal is proportional to $f^2_{\rm rot}$, the resulting waveform will be a sort of inverse chirp. 
The amplitude is also directly proportional to the remnant deformations with respect to axisymmetry around the spin axis, encoded by a quadrupolar ellipticity. Significant deformations can be expected, e.g., if the obejct has a strong internal magnetic field (e.g., \citealt{Cutler2002,DallOsso2009,Ciolfi2013, DallOsso2018, Lander2020}). 

\section{Electromagnetic Emission}
\label{sec:EM}

\begin{figure}
\centering
\includegraphics[width=0.7\columnwidth]{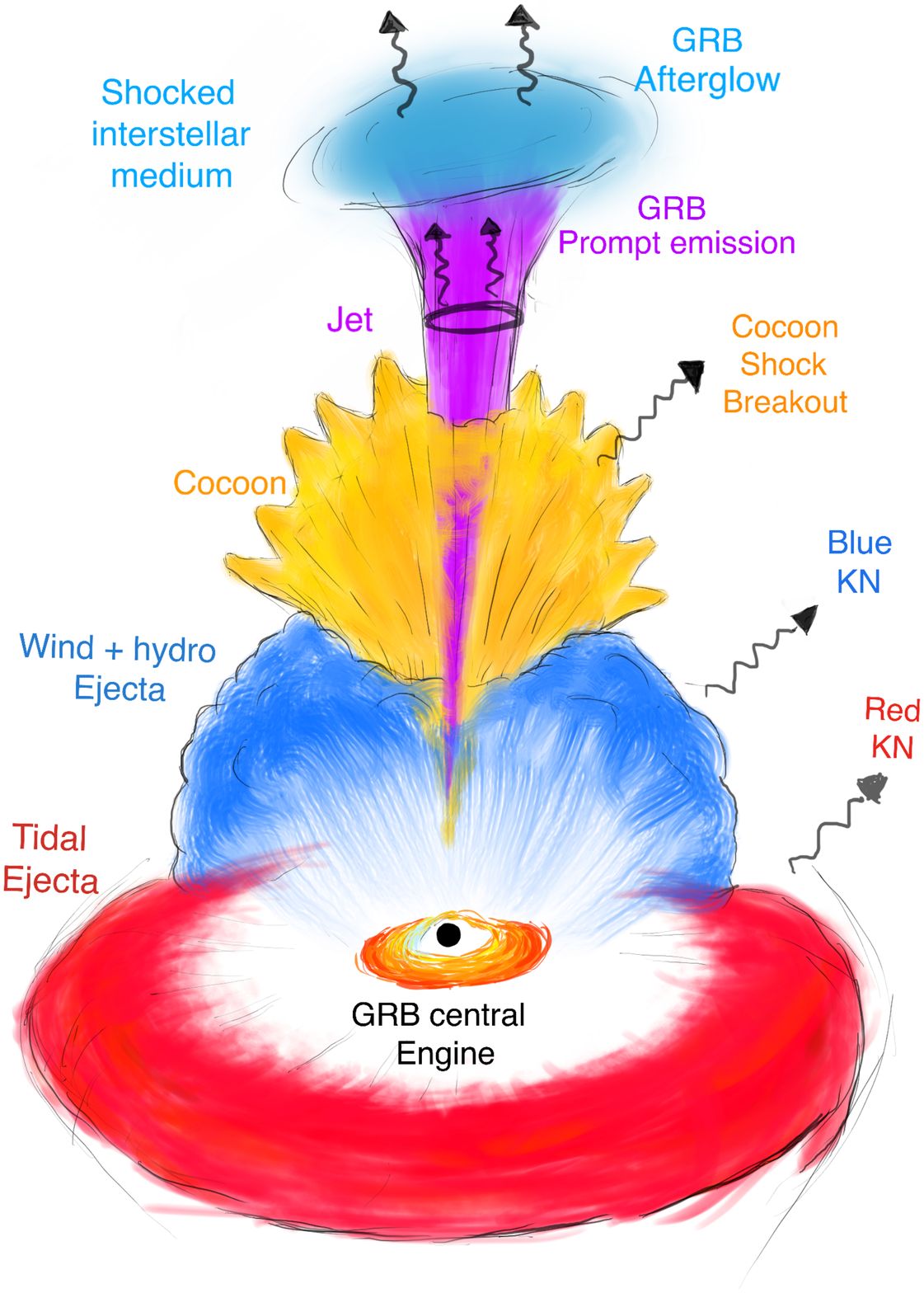}
\caption{Artistic representation of the scenario following a \bns/\nsbh merger, when an accreting \ac{BH} is formed. The red component denotes the tidal ejecta, the blue component the hydrodynamic and wind ejecta, the purple component the jet and the yellow component the matter of the ejecta heated by the jet (cocoon). The different components are not represented in scale.}
\label{fig:EMgeometry}
\end{figure}

In general, when at least one \ac{NS} is involved in the merger, emission of photons is expected along with \acp{GW}. In Fig.~\ref{fig:EMgeometry} we illustrate the different EM emission components represented with a different color.
The only cases where the \ac{EM} emission may be suppressed is when a \ac{BH} is promptly formed and leaves no accretion disk nor ejected material in NS binary mergers, and when the \ac{NS} is swallowed by the \ac{BH} without being disrupted in \nsbh mergers.

It is also possible that \ac{EM} emission takes place even before the merger. This kind of emission, known as \emph{precursor emission}, would occur irrespective of the final fate of the merger, namely it would be present even if a \ac{BH} with no accretion disk results from the coalescence. Precursor emission will be discussed in Sec. \ref{sec:precursor}.

The remnant object resulting from the merger, either a differentially rotating massive \ac{NS} or a \ac{BH} surrounded by an accretion disk, is expected to launch a jet of relativistic matter\footnote{This is commonly observed in other astrophysical sources consisting in accreting \acp{BH}, such as active galactic nuclei \citep[][and references therein]{Blandford2019} and microquasars \citep[][and references therein]{Mirabel1999}.} (purple component in Fig.~\ref{fig:EMgeometry}) by physical processes which are, at present, not completely understood. The mechanisms to launch a jet are discussed later in Section~\ref{sec:GRB_centralengine}. 

Once formed, the jet drills through a dense circum-burst medium, constituted by the material previously expelled by the merger (blue component in Fig.\ref{fig:EMgeometry}). During the jet crossing, the material in front of the jet's head is heated and moved aside, forming an hot structure called \emph{cocoon} (yellow component in Fig. \ref{fig:EMgeometry}). The cocoon, in turn, exerts a transverse pressure which confines and further collimates the jet \citep{Bromberg2011, Lazzati2019, Salafia2019}.  
If sufficiently energetic, the jet successfully emerges from the surrounding material (whose radius at this time is of order $\sim 10^{10}\,\rm cm$), otherwise the jet can be choked. 
In the successful case, the jet travels until it becomes transparent at a radius of order $\sim 10^{12}\,\rm cm$ from the central engine \citep{Piran1999, DaigneMochovitch2002}.
At a distance range of $10^{13}-10^{16}\,\rm cm$ the jet dissipates part of its kinetic energy powering an energetic gamma-ray radiation known as the \emph{\ac{GRB} prompt emission}. Alternatively, this emission can occur at the photosphere if the energy is dissipated below it.   
The internal dissipation mechanism is still poorly understood. Later on, the jet decelerates shocking the interstellar medium (in light blue in Fig. \ref{fig:EMgeometry}), powering a fading synchrotron emission from X-ray to radio called \emph{\ac{GRB} afterglow}. 
 \ac{GRB} prompt and afterglow emission will be discussed in more detail in Section \ref{sec:SGRB}. It is worth noticing that further \ac{EM} radiation can originate from the cocoon when it breaks out from the circum-burst medium and when it cools down by thermal emission. This kind of counterpart will not be discussed in the present review and we refer the interested reader to \citet{Nakar2019}.

The other component of the matter left outside after the \ac{CBM} and unbound from the central remnant (blue and red components in Figure \ref{fig:EMgeometry}) is rich of free neutrons and neutron-rich nuclei and represents an ideal site for the r-process nucleosynthesis of heavy elements \citep{LattimerSchramm1974, Symbalisty1982}, whose radioactive decay heats up the material itself. This results in a thermal emission from the ejecta that generates a transient known as \emph{\ac{KN}} which will be discussed in more detail in Section \ref{sec:kilonova}. It is also worth noting that the shock onto the interstellar medium of these mildly relativistic and sub-relativistic outflows could in principle power a synchrotron emission expected to be observable in radio frequencies \citep{NakarPiran2011}.

If the merger remnant is a very long-lived \ac{SMNS} or a stable \ac{NS}, the conditions to form a magnetosphere can be met, leading to a magnetic dipole spindown radiation phase. 
Such emission would then energize the surrounding material (i.e.~the previously emitted baryon rich ejecta) powering an additional \emph{\ac{SDPT}}. 

The \ac{EM} counterparts of \ac{CBM} briefly described here will be discussed in more detail in the following, with a special focus on the open problems in the field.

\subsection{Precursor Emission}
\label{sec:precursor}

The precursor emission is an \ac{EM} signal preceding in time the peak of \ac{GW} radiation, namely it is supposed to be emitted during the inspiral phase of the binary.

The general idea is that the signal is generated by the interaction between \ac{NS}'s magnetospheres (in case of a \bns system) or the orbital motion of a non/weakly-magnetized object through the magnetosphere of the companion. The latter framework, known as \emph{unipolar inductor} has been studied analytically both for \bns \citep{V96, HansenLyutikov2001, Piro2012, L12, Sridhar2020} and \nsbh systems \citep{McWilliamsLevin2011}. In this framework the motion of the non-magnetized object through the companion magnetic field generates, due to Faraday's law, an electromotive force on the non-magnetized object. In this way a DC circuit is established between the two stars, where the field lines behave like wires and the non-magnetized object as a battery. The accelerated charges may dissipate on the stellar surface or in the space between the two stars. In the latter case they may emit radio waves in a pulsar like fashion or trigger a pair cascade resulting in a wind emitting in the X-rays \citep{HansenLyutikov2001, Piro2012}.   
Furthermore, shocks arising within the wind may generate an observable coherent radio emission through syncrothron maser process \citep{Sridhar2020}.

The problem has been studied also with general relativistic resistive \ac{MHD} \citep{P13}, special relativistic \citep{Sridhar2020}, force-free \citep{MostPhilippov2020} and particles-in-cell \citep{Crinquand2019} simulations. These works allowed to appreciate the important role of magnetic reconnection in accelerating particles, which can further contribute to the emission

Since this kind of emission takes place during the inspiral phase, its occurrence is insensitive of the merger product, namely it can be generated even in those cases in which no matter is left outside the newly formed \ac{BH}. In these cases precursors represent the only possible \ac{EM} counterpart. Although there is not a compelling observation for the above described precursor yet, surveys able to detect them can provide candidate counterparts to be used for a coincident search of \acp{GW}.

\subsection{ $\gamma$-ray bursts}
\label{sec:SGRB}

\begin{figure}
    \centering
    \includegraphics[width = 0.45
    \columnwidth]{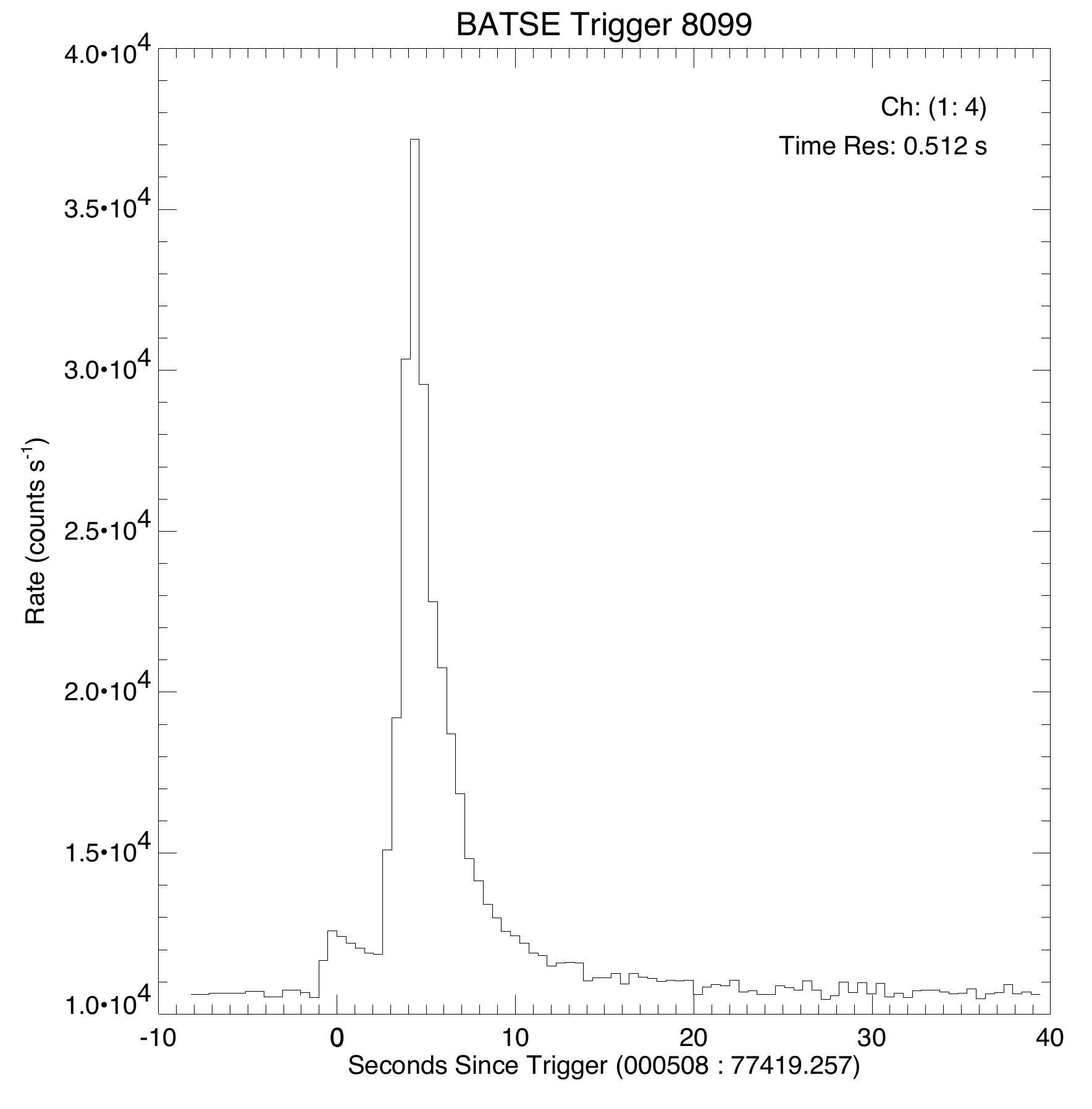}
    \includegraphics[width = 0.45
    \columnwidth]{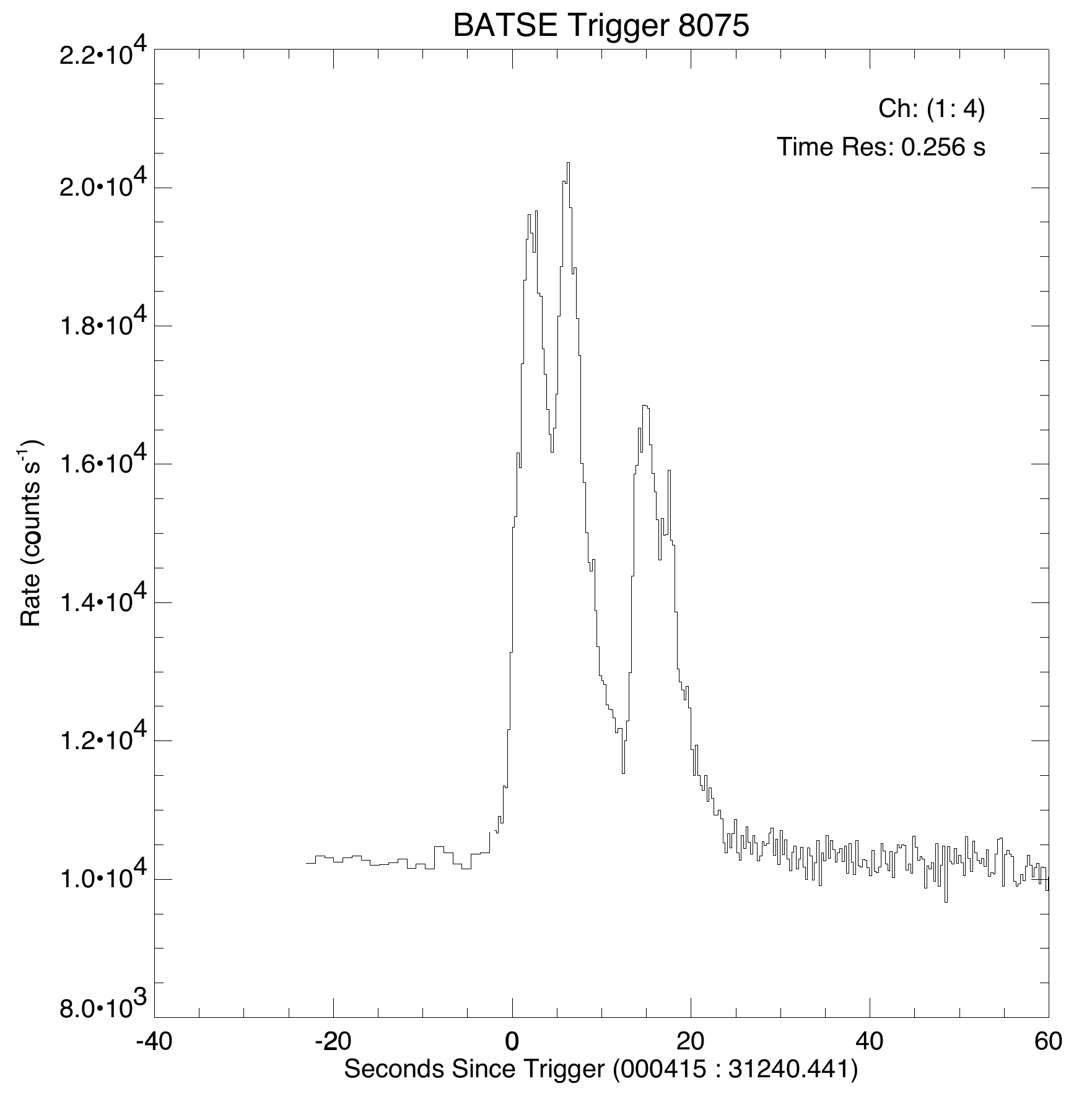}
    \includegraphics[width = 0.45
    \columnwidth]{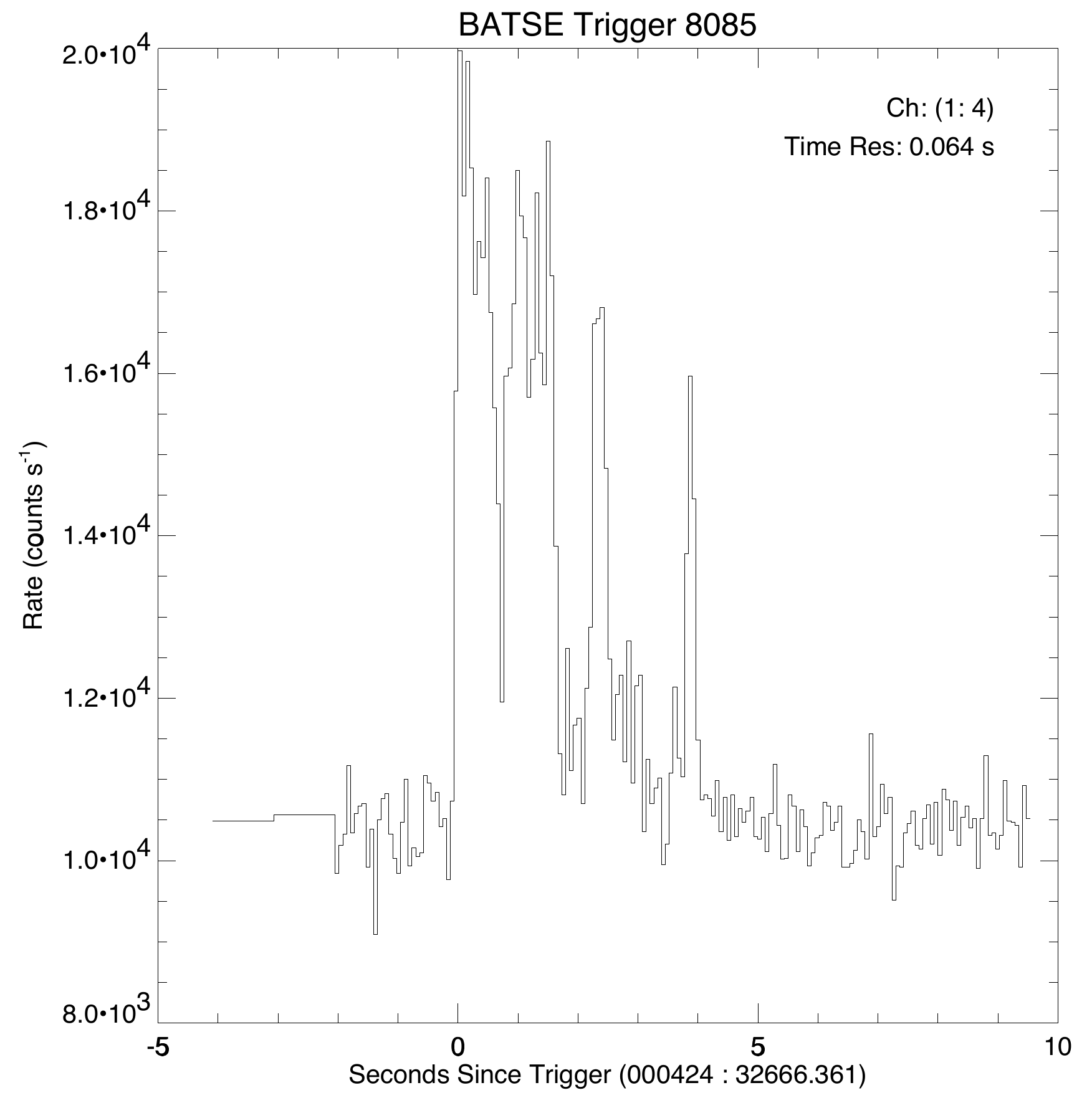}
    \includegraphics[width = 0.45 \columnwidth]{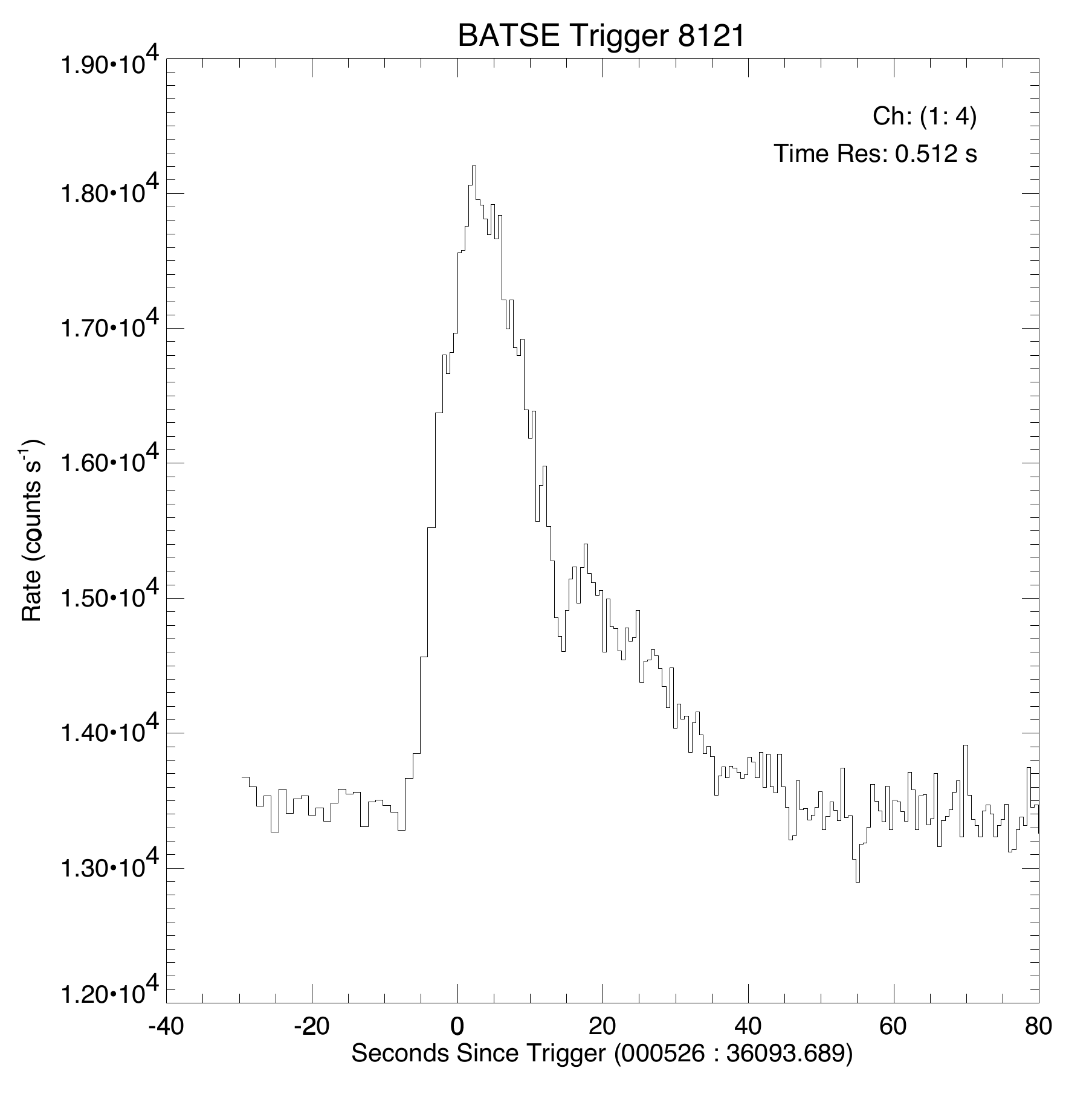}
    \caption{Examples of \ac{GRB} prompt emission lightcurves ($E>20\, \rm keV$) from the online BATSE catalog: https://gammaray.nsstc.nasa.gov/batse/grb/lightcurve/
    }
    \label{fig:prompt_LC}
\end{figure}

\begin{figure}
    \centering
    \includegraphics[width = \columnwidth]{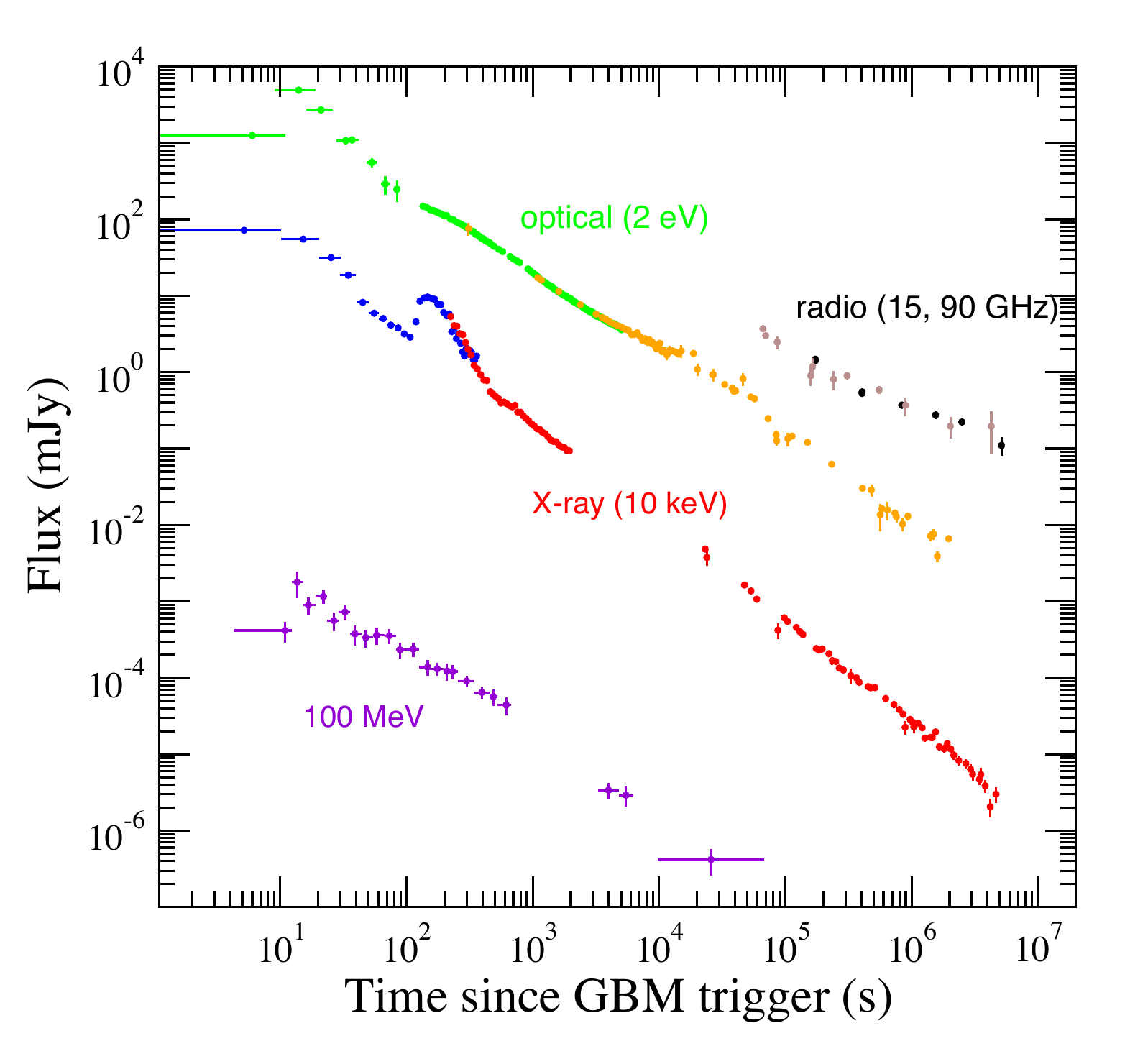}
    \caption{Example of a GRB afterglow in different spectral ranges. The data reported are for GRB 130427A afterglow \citep{Painatescu2013}. (Figure courtesy of Alin Panaitescu).}
    \label{fig:afterglow_LC}
\end{figure}

\acp{GRB} are highly energetic transient astrophysical sources of extragalactic  origin, which have non-thermal spectra peaking at $10$ keV - $10$ MeV. The $\gamma$-ray flaring activity of \acp{GRB}, which is called \emph{prompt emission}, lasts typically less than hundred of seconds and its lightcurve manifests with a plethora of different morphologies and a time variability (timescale in which the fluxes vary by more than a factor of 2) that can be of the order of few $\rm ms$ \citep{Walker2000}. Examples of typical prompt emission lightcurves are shown in Figure \ref{fig:prompt_LC}, where the \ac{GRB} data are taken by \ac{BATSE} onboard the \ac{CGRO} and the lightcurves are expressed in terms of counts per seconds.
The prompt emission is followed by the fainter \emph{afterglow} across all \ac{EM} spectrum which decays in time. An example of a typical afterglow lightcurve in different spectral ranges is showed in Fig. \ref{fig:afterglow_LC}.

The duration of the prompt emission shows a bimodal distribution \citep{Kouveliotou1993}, with two peaks at $0.3\,\rm s$ and $30\,\rm s$, which unveils the presence of two distinct \ac{GRB} populations, also characterized by a different hardness of the spectrum: the population of short duration events with a harder spectrum, called \acp{SGRB}, and the population of long duration events with softer-spectrum, called \acp{LGRB}. The duration commonly used to separate short \acp{GRB} from long ones is $2\,\rm s$ in the \ac{BATSE} energy range ($20 - 600\,\rm keV$).

From the $\gamma$-ray fluence and the distance of the source it is possible to calculate the isotropic equivalent energy of the burst, which can reach $E_{\gamma, \rm ISO} \sim 10^{54}\,\mathrm{erg}$ \citep{Briggs1999,Abdo2009} for the most energetic \acp{GRB}. However, there are several very strong  evidences that the emission is not isotropic but beamed \citep{Rhoads1997,Rhoads1999,Frail2001}. This reduces the value of the real energy by a factor equal to the solid angle of collimation $\theta^2/2$, limiting it in the range $10^{49}-10^{51}\,\mathrm{erg}$ \citep{Frail2001,Panaitescu2001,Bloom2003}. 

The variability timescales $\delta t$ can give us further insights on the origin of this emission. It constrains the emitting region to a characteristic size of $l \sim c \delta t$. This size together with the high luminosity and the fact that the emission lies in $\gamma$-ray frequency range brings to a source that should be so compact to be opaque to $\gamma -\gamma$ pair-production, which is in tension with the non-thermal nature of the observed spectra. This problem, known as \emph{compactness problem} \citep{Ruderman1975,Goodman1986,Paczynski1986,Piran1995}, is solved by requiring that the source is moving relativistically towards the observer with a Lorentz factor $\Gamma \ge 100$. This results in an increase of the true size of the source by a factor $\Gamma^2$, and in a blue-shift of the energy, which increases in the observer frame by a factor $\Gamma$ (for a more detailed discussion of the compactness problem see \citealt{P05}).

All the above considerations indicate that \acp{GRB} are generated by a collimated source moving relativistically towards the observer, the relativistic jet. The jet must be launched by some physical processes occurring within a short living\footnote{Since \acp{GRB} does not repeat.} object, which is usually referred to as \emph{central engine}, originated from a \emph{progenitor} system.

\acp{SGRB} and \acp{LGRB} have different progenitors.
This was revealed by studies of the GRB energetics and redshift distribution\citep{Nakar2007}, properties of their afterglow \citep{Panaitescu2001b,Kann2011}, 
their host galaxies\citep{Berger2009,Fong2013} and the recent observations of \acp{GW}.

A fraction of \acp{LGRB} optical afterglows show a late-time (in few weeks) bump due to an emergent associated type Ic supernova\footnote{Type Ic supernovae are supernovae whose spectrum is lacking of hydrogen lines (which are present in type II supernovae), silicium absorption line (present in type Ia supernovae) and helium lines (present in in type Ib supernovae).  They are believed to be produced by core-collapse of massive stars which lose their hydrogen and helium envelopes prior to the explosion \citep{Woosley2006}}. 
Thus, the \ac{LGRB} afterglow observations empirically associates the sources of \acp{LGRB} with the death of massive stars (masses $M>20M_{\odot}$) \citep[][and references therein]{Woosley2006}. 
The same conclusion comes from the study of their host galaxies which are typically irregular and star forming galaxies \citep{Fruchter2006}. \acp{LGRB} are preferentially located in the brightest spots of their host galaxies, i.e.  highly star forming regions where the massive stars are born. The supernovae signatures in \acp{LGRB} observations and the preferential location in actively star forming regions, are strong evidence that massive stars are the progenitors of \acp{LGRB}. 

On the other hand, most of the host galaxies of \acp{SGRB} show a relatively low star formation rate \citep{Fong2010}. When \acp{SGRB} are in star-forming galaxies, they are located in regions with lower star formation rate and associate with an older star population with respect to \acp{LGRB}. This indicates that progenitors of \acp{SGRB} are older than those of \acp{LGRB}, since the region where they had formed is not active anymore or/and they had enough time to move away from their birth place. Astrophysical events satisfying all the required characteristics to be progenitors of \acp{SGRB} are the mergers of \bns or \nsbh binaries \citep{Blinnikov1984, Paczynski1986, ELPS89}.
The first strong observational evidence to identify their progenitors came from the coincident detection of the first \ac{GW} signal from the merger of \bns, GW170817,
by the Advanced LIGO and Virgo interferometers and the \ac{SGRB} named GRB\,170817A by the $\gamma$-ray detectors FERMI-GBM and INTEGRAL \citep{GW170817MMA, Goldstein2017, Savchenko2017}. Together with the multi-wavelength observations spanning more than two years, which demonstrated the presence of a relativistic jet, this proved that the merger of two NS (in this specific case of total mass of $2.74_{-0.01}^{+0.04}M_{\odot}$) can give rise to \acp{SGRB}.

Both core-collapse of a massive star and \ac{CBM} involving \ac{NS} are able to generate \acp{GRB} by forming either a temporary accretion disk around the remnant \ac{BH} or a massive (meta)stable \ac{NS}. These remnant systems are potentially able to power a relativistic jet via different mechanisms, and thus they are considered the most plausible \ac{GRB} central engines.
The internal dissipation of the jet energy at $\rm R_{\gamma} \sim 10^{13} \, cm - 10^{16}\, cm$ produces the observed short-duration ($\rm 10^{-3}-10^{3}\,s$) prompt emission seen in the keV-MeV energy range \citep{Rees1994,Meszaros1994,Daigne1998,Piran1999} of, alternatively, the jet energy can be dissipated below the photosphere ($R \sim 10^{12}\,\rm cm$) and released at the photosphere (e.g. \citealt{Rees2005,Peer2008, Beloborodov2010,Lazzati2013,Bhattacharya2020}).

\begin{figure}
\centering
\includegraphics[scale=0.3]{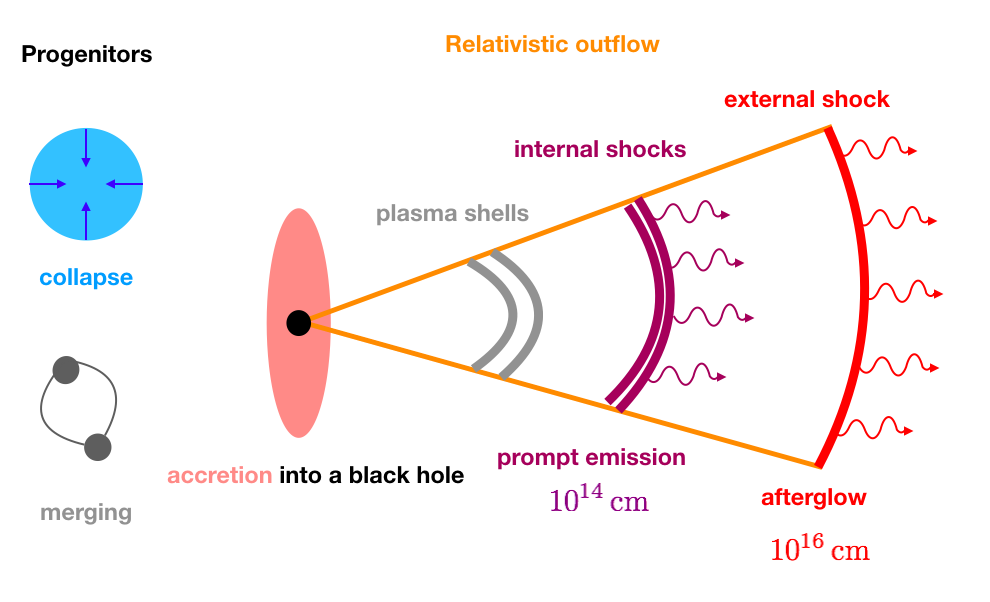}
\caption{Pictorial representation of the basic model of the GRB: the core-collapse of a massive star or the coalescence of a \bns (or \nsbh) leads to the formation of an accreting \ac{BH}, which launches an ultra-relativistic outflow in a form of a jet. Internal dissipation of the jet's kinetic energy through shocks produces the prompt emission in the keV-MeV range. The forward shock of the jet with ambient medium forms the afterglow radiation observed in the X-ray, optical and radio bands.}
\label{fig:sketchGRB}
\end{figure}

The deceleration of the jet in the surrounding medium gives rise to the long-lived afterglow observed in the X-ray, optical, and radio bands\citep{Paczynski1993,Meszaros1997,Sari1998}. The characteristic physical size of the afterglow emitting region is $\rm R_{aft} \sim 10^{16}-10^{17} \, cm$. The sketch of the basic model of the GRB phenomenon is shown in Fig. \ref{fig:sketchGRB}. 

Since \acp{SGRB} originates from \acp{CBM}, they are the main objective of the present review. 
Nevertheless, various aspects discussed in the following apply also to the case of \acp{LGRB}.

\subsubsection{Central engine of short GRBs}
\label{sec:GRB_centralengine}
One of the fundamental unresolved question in the \ac{GRB} physics lies in the identification of the powerhouse of the relativistic jet (see, e.g., \citealt{Ciolfi2018} for a more extensive review). The proposed scenarios as central engine of \acp{GRB} needs to satisfy some  minimal physical requirements.
 
One of them is the ability to power a jet with bulk Lorentz factors of order $\Gamma \ge 100$. This implies that the formed jet should contain a small amount of baryons otherwise their large inertia will prevent the fast motion \citep{ShemiPiran1990}. 
Moreover, highly variable light curves of the prompt emission suggest a discontinuous release of the energy. As mentioned before, the main candidate that satisfies the above requirements is the hyper-accretion of a stellar mass BH, accreting at a rate $\dot{M} = 1\,\rm M_\odot/s$ \citep{Woosley1993}.
This rate is very high and exceeds by several orders of magnitude the Eddington limit, above which the feedback of radiation released by the accretion is strong enough to sweep off the infalling material hampering in this way the accretion. This limit in \acp{GRB} can be overcome because the disks are thought to be in a peculiar accretion regime, named \emph{neutrino-dominated accretion flow}, where due to the high density and pressure photons are trapped inside the matter and the neutrino emission dominates the disk cooling \citep[see][for a recent review]{Liu2017}.
Alternatively, a magnetar in fast rotation (periods of milliseconds) can power the relativistic jet via its rotational energy \citep{Usov1992}. 

The duration of the prompt emission is a further constrain on the nature of the central engine. 
While the prompt emission lasts for less than 2 seconds, we can observe late-time X-ray plateaus and/or flares (at $10^{3}-10^{4}$ s) \citep{Nousek2006,Zhang2006,OBrien2006,Chincarini2007,Falcone2007}. If we
interpret these as a continuation of the energy output from the central engine \citep{Rees1998}, then the central engine needs to be very long-lived and/or reactivated at such late times (e.g. \citealt{Dai1998,Zhang2001,Yu2010}).

In the case of an accreting \ac{BH} central engine, the latter can power the jet by different mechanisms. The most discussed are the neutrino-antineutrino annihilation process \citep{Goodman1987, ELPS89, ZB11} and the \ac{BZ} mechanism \citep{BZ77}. 
The first mechanism is based on the fact that in the hottest inner part of the accretion disk neutrino cooling is expected to become very efficient. When this happens, the annihilation of neutrino and antineutrino pairs above the disk should lead to the formation of an expanding fireball (electron-positron plasma mixed with photons). An estimate of the expected luminosity of the jet in this scenario comes from the power of neutrino-antineutrino annihilation and it is $\sim 10^{52}$ erg/s for a $3\,M_\odot$ black hole with an accretion rate $\dot{M} \sim 1 M_{\odot} s^{-1}$ \citep{ZB11}. It is worth noticing that, in contrast, photon-photon pair production is not an available mechanism because of the above-mentioned regime of \ac{GRB}'s disks, in which photons are trapped and only neutrinos and antineutrinos are able to escape.
The second mechanism, currently favoured, consists in extracting the rotational energy of a Kerr \ac{BH} through a strong magnetic field threading it, which is sustained by electric currents within the accretion disk. 

The energy is transported away in the form of a Poyinting flux flowing mainly along the \ac{BH}'s rotational axis. For such a Poynting flux to exist a non-zero toroidal component of the magnetic field is required along the axis\footnote{The Poynting flux is $\mathbf{S} \propto \mathbf{E} \times \mathbf{H}$, where $\mathbf{E}$ and $\mathbf{H}$ are the electric and auxiliary magnetic fields. Faraday's law along with axysimmetry and stationarity guarantees that the electric field is purely poloidal. So to have a non-vanishing radial component of $\mathbf{S}$ along the axis a toroidal component of the magnetic field is required.}. The toroidal magnetic field is generated by a poloidal current, streaming along the poloidal field lines that behave as wires. The electric currents must be sustained by an electromotive force. This is provided by a purely general relativistic effect, namely the frame-dragging around a rotating mass. The frame-dragging applied to the poloidal magnetic field lines results in the appearance of an electric field providing the required electromotive force. It is possible to prove that inside the \ac{BH}'s \emph{ergosphere} - a region outside the event horizon where the frame-dragging is so intense that even photons are dragged to co-rotate with the \ac{BH}-  in order to attain the screening of this electric field, the required above mentioned poloidal current cannot vanish \citep{Komissarov2004}. 

A rapidly rotating \ac{NS} remnant, with a period of the order of few milliseconds and dipolar magnetic field strenghts of order $\sim 10^{15}$ G represents a viable alternative to the accreting \ac{BH} as a machine for the production of a relativistic jet. This central engine is often called a \emph{millisecond magnetar}. 
Note that this type of engine can only apply to the case of \bns mergers. 
In such a scenario, the NS remnant, characterized by strong differential rotation, gradually builds up a strong helical magnetic field structure along the spin axis and the corresponding magnetic pressure gradients accelerate a collimated outflow in such direction \citep{Ciolfi2020a}.
This magnetorotational mechanism will remain active only as long as the differential rotation in the core of the NS remnant persists, which is typically limited to sub-second timescales. 

Recent simulations showed that the neutrino-antineutrino annihilation mechanism is likely not powerful enough to explain \ac{SGRB} energetics (\citealt{Just2016,Perego2017} but see \citealt{Salafia2020}), pointing in favour of the alternative magnetically driven jet formation.
A number of numerical relativity simulations of \bns and \nsbh mergers including magnetic fields as a key ingredient studied the potential to form a \ac{SGRB} jet in the accreting \ac{BH} scenario (e.g., \citealt{Rezzolla2011,Paschalidis2015,Kiuchi2014,Kawamura2016,Ruiz2016}) and in the magnetar scenario (only for NS-NS mergers; e.g., \citealt{CiolfiKastaun2017,Ciolfi2019,Ciolfi2020a}). While a final answer is still missing, the most recent results support the \ac{BH} scenario, either by showing that an incipient jet could form in this case \citep{Paschalidis2015,Ruiz2016} or by showing that the collimated outflow from a magnetar would most likely be insufficient to explain the energy and Lorentz factors of \acp{SGRB} \citep{Ciolfi2020a}\footnote{Recently, \citet{Mosta2020} reported the formation of a mildly relativistic collimated outflow from a massive NS remnant, also showing that the presence of neutrino radiation can lead to a more powerful outflow compared to the case without neutrino radiation. In their set of simulations, however, a strong ($B = 10^{15}\, \rm G$) dipolar magnetic field is superimposed by hand on the NS remnant at 17ms after merger. This assumption, while having a major impact on the result, may not be realistic, which casts doubts on the conclusion that massive NS remnants would be able to launch powerful jets.}. 

\subsubsection{Jet acceleration}

Once the jet is formed, it must accelerate to ultra-relativistic velocity. How it happens is strongly connected to the nature of the jet. There exists two limiting cases: a hot internal-energy-dominated jet and a cold magnetic-energy-dominated jet. 

The first case is the most explored scenario and is known as \emph{hot fireball} model \citep{CavalloRees1978, Paczynski1986,Goodman1986}. The general idea beyond this framework is that the internal energy of the jet is converted into bulk kinetic energy, such that the temperature of the fireball decreases while the jet accelerates.
If we use the typical observed prompt emission luminosity $L \sim 10^{52}$ erg/s and the variability time-scale of $10^{-2}$ s, we can estimate the initial temperature of the ejecta as $T \sim \left( L/4 \pi (c \delta t)^2 \sigma_{B} \right)^{1/4} \sim 2 \times 10^{10} K$. At this high temperature photons are coupled with electron-positron pairs and baryons. Jets with small enough amount of baryons will undergo an adiabatic expansion while photons cool. 
Therefore, the initial energy of photons, electrons, and positrons is transferred to protons. The acceleration of the jet continues until photons de-couple from baryons \citep{ShemiPiran1990, Piran1999}. 

The photosphere can be defined as a surface at which the optical depth to the Thomson scattering ($\tau$) is  equal to 1. The optical depth is $\tau \sim n_{p}\sigma_{T} R/2 \Gamma$, where $R$ is the radius of the ejecta and $n_{p}$ is the proton density in the outflow. The comoving proton density estimate comes by the proton flux in the jet $\dot{M}$ as  $n_{p}' = \dot{M}/(4 \pi R m_{p} c \Gamma)$. Defining the ratio between the radiation luminosity $L$ and the baryonic load $\dot{M}$  as $\eta = L/\dot{M}c^{2}$ we can get the photosphere radius $R_{ph} = R(\tau=1)$ as $R=L \sigma_{T}/8 \pi m_{p} c^{3} \eta \Gamma^{2} $. Its typical value is $R_{ph} \sim 6 \times 10^{12}$ cm (if we use $L \sim 10^{52}$ erg/s, $\eta \sim 100$ and $\Gamma \sim 100$) \citep{Piran1999,DaigneMochovitch2002}. The parameter $\eta$ is the test parameter for identification of the jet composition. If thermal emission from the fireball is observed in the prompt emission spectra, its temperature constrains $\eta$ and we have knowledge of  the initial form of the jet's energy. Unfortunately, we do not have significant observational claims on the thermal components in \acp{GRB} spectra to answer to this question.

The alternative is a cold jet, dominated by the Poynting flux (as, e.g., in the \ac{BZ} mechanism).
These kind of jets can be accelerated by converting the magnetic energy in kinetic energy of the bulk flow. This can occur, for example, due to the adiabatic expansion of the outflow, where, in the axisymmetric and stationary limit, the conservation of mass and energy flux leads to identify the following conserved quantity: 
\begin{equation}
    \Gamma(1+\sigma) = \rm constant,
\end{equation}
where $\sigma$ is the plasma magnetization, defined as the ratio of Poynting flux and matter energy flux. This means that the magnetic energy stored in the jet is able, in principle, to accelerate the jet up to the maximum Lorentz factor of $\Gamma_{\rm max} = 1 + \sigma_0 \sim \sigma_0$, where $\sigma_0 \gg 1$ is the magnetization at the base of the jet. The acceleration is due to magnetic pressure gradients associated with the build-up of toroidal magnetic field via the winding of field lines caused by the fast and differentially rotating central engine. 
However, in order to reach $\Gamma_{\rm max}$, the head of the jet must be in causal contact with its base and this condition is reached as long as the flow is subsonic with respect to the speed of a fast-magnetosonic wave. Since this speed is approximately equal to the (relativistic) Alfvén velocity, whose corresponding Lorentz factor is $\Gamma_A = \sqrt{1+\sigma}$, the condition $\Gamma = \Gamma_A$ sets a limit to the Lorentz factor of $\Gamma_{\rm MS} = (1+\sigma_0)^{1/3}\sim \sigma^{1/3}$, reached at the distance $R_{\rm MS}$ from the central engine, which is known as \emph{magnetosonic point} \citep{GoldreichJulian1970}. This value can be increased by a factor of $\theta^{-2/3}_j$, when the outflow is collimated within an angle $\theta_j$ due to the confinement exerted by the circumburst medium (the envelope of the progenitor star for \acp{LGRB}, the merger ejecta for \acp{SGRB}) \citep{Tchekhovskoy2009}.  

Above the magnetosonic point, the jet can be accelerated both due to adiabatic acceleration, when a non-stationary outflow in place of a time-independent jet is considered \citep{Granot2011}, or by the dissipation of magnetic field due to magnetic reconnection, which converts magnetic energy into thermal energy and, in turn, into bulk kinetic energy, as in the hot fireball case \citep{Drenkhahn2002, DrenkhahnSpruit2002}. It is worth noting that dissipation processes able to convert a Poynting flux dominated jet into a hot fireball can occur also where the interaction between the jet and the circumburst medium leads to the jet collimation, due to magnetic reconnection driven by unstable internal kink modes as appears in 3D \ac{MHD} simulations \citep{Bromberg2016} 

For a more detailed description of the jet acceleration mechanisms we refer the reader to \citet{KuZh2015} and \citet{ZhangBook}.

\subsubsection{Jet dissipation mechanisms}

The observed prompt emission is originated by the dissipation of the jet's energy. The short variability of prompt emission lightcurves suggests that the dissipation occurs within the jet (internal dissipation) because the lightcurve produced by the forward shock of the jet into the surrounding ambient medium (external dissipation) is expected to be smoothed over a timescale which is  longer than the total duration of the burst (see \citealt{Fenimore1996,Sari1997,Piran1999, P05}).

The internal shocks model is widely used as a dissipation mechanism in baryon dominated jets \citep{Rees1994}. In that model the central engine produces outflows with random Lorentz factors. The faster part of the outflow catches the slower one. 
The shocks propagate in both shells converting the bulk kinetic energy of the flow into kinetic energy of the particle accelerating them, and also leading to a local magnetic field amplification (\citealt{Heavens1988,Sironi2011} for a review on the acceleration of particle in shocks we refer the interested reader to \citealt{shockreview}).
Relativistic electrons in magnetic field are expected to radiate their energy through synchrotron and inverse Compton processes. 

The main advantage of the internal shocks model is the possibility to have short variability time-scale.
For simplicity, let's consider that the central engine produces two shells (with bulk Lorentz factors $\Gamma_1$ and $\Gamma_2$) with time difference $\delta T$. If the first shell is slower $\Gamma_1 < \Gamma_2$ then they collide at time $t_{coll}$ defined by $v_1 t_{coll} = v_2 (t_{coll}-\delta T)$ (where $v_1$ and $v_2$ are shells' velocities). The radius at which shells collide is $R_{coll} = t_{coll} v_1 \sim 2 c \Gamma_{1} \delta T \kappa$ where $\kappa = \Gamma_2/\Gamma_1$. The observed variability tracks then the intrinsic central engine variability. However, the efficiency of internal shocks dissipation, which depends on the velocity difference between the two shells, is very low (about $20 \%$ or lower) \citep{Kobayashi1997,DaigneMochovitch2002} and it represents one of the main issues of this model (but see \citealt{Beniamini2016}). Moreover, it was realized that the magnetization of the jet should be relatively low, i.e. $\sigma \sim 0.01-0.1$, otherwise the shocks will be suppressed \citep{Sironi2015b}.  

Alternatively, the prompt emission can be released at the photosphere, from sub-photospheric dissipation (e.g. \citealt{Rees2005,Peer2008, Beloborodov2010,Lazzati2013,Bhattacharya2020}).
The sub-photospheric dissipation can proceed through different mechanisms. One example is radiation dominated shocks, where the dissipation is controlled by photon scattering (\emph{e.g.} \citealt{Beloborodov2017}; see also \citealt{LevinsonNakar2020} for a recent review about the topic). Another mechanism involves the turbulence within the jet, which transfer the kinetic energy of the flow to plasma particles, if the turbulence cascade reaches the microscopic scales, or directly to the radiation, in case turbulence cascade is damped at larger scales by bulk Compton scattering \citep{Zrake2019}. Furthermore, if free neutrons are present, nuclear collision can constitute another important dissipation channel \citep{Beloborodov2017}.

In the case of a Poynting flux dominated jet the basic dissipation mechanism, as already mentioned in the previous section, is the magnetic reconnection. 
During the reconnection the magnetic energy is dissipated to accelerate the charges, which then radiate through different emission processes (\emph{e.g.} synchrotron emission). Particle acceleration can proceed also through Fermi mechanism. This occurs when the particle scatters through the magnetic islands produced due to tearing instabilities in reconnection events, and moving close to the Alfvén speed (for acceleration of particles in reconnection layers see \citealt{Spruit2001,DrenkhahnSpruit2002,Sironi2014,Sironi2015b,Petropoulou2019}). Moreover, tearing instability is also fundamental for enhancing the rate of reconnection events (for reviews on tearing instability see \emph{e.g.} \citealt{Loureiro2012, Comisso2016}).

In \acp{GRB} magnetic reconnection can be triggered by internal shocks, as proposed by \citet{ZhangYan2011} in their Internal-collision-induced Magnetic Reconnection and Turbulence (ICMART) model. In the ICMART model the shock between shells endowed by ordered magnetic field (generated by the winding of magnetic field line in proximity of the central engine) cause a tangling of the field line, which favour magnetic reconnection on a length scale much smaller than the transverse size of the jet. Moreover, during reconnection process, the plasma is accelerated and this further distort the field lines, triggering further reconnection processes in a cascade fashion. The charges accelerated toward the observer radiates through synchrotron process, and this should generate the $\gamma$-ray radiation constituting the \ac{GRB} prompt emission. The distance from the central engine at which the dissipation occurs in \citet{ZhangYan2011} model is rather large, of the order of $\sim 10^{16}\,\rm cm$, however the millisecond observed variability can be attained considering that not the entire jet, but smalls regions of it, switch-on independently\footnote{In \ac{GRB} models this feature is often refer to as "mini-jet" \citep{KumarPiran2000, Narayan2009}}. Contrary to the internal shock model described at the beginning of this section, the ICMART model has a very high radiative efficiency.

\subsubsection{Afterglow}

As mentioned before, the deceleration of the jet in the surrounding medium gives rise to a long-lived emission, called \emph{afterglow}, observed in radio frequencies, optical, X-ray, GeV and very recently also at TeV energies \citep{MAGICGRB2019, HESSGRB2019}. 
The characteristic radius, measured from the central engine, at which the deceleration, and thus the afterglow, occurs is $\rm R_{aft} \sim 10^{16}-10^{17} \, cm$.

The afterglow theory was developed before the first afterglow had been observed. It predicted an emission with a temporal power-law fading $ t^{-\alpha}$ in a wide range of frequencies after the deceleration time (define as the time at which the jet Lorentz factor halves). Here $\alpha$ depends on the power-law index $p$ of the distribution of the emitting charges (see discussion below) and $\alpha \sim 1$ for expected values of $p$ \citep{Paczynski1993, MeszarosRees1997, Sari1998}. This prediction has been confirmed by the first afterglow observations (mainly in the optical band)\footnote{Actually, the first afterglows have been observed in X-rays, but the observations occurred at very late time from the \ac{GRB} prompt, due to the large uncertainty $\gamma$-ray instrument in localizing the source along with the small field of view and the large time of re-pointing of X-ray telescopes.}. 
However, later observations with the Neil Gehrels Swift Observatory, (hereafter \emph{Swift}), show that the X-ray lightcurve departures from the simple power-law behaviour , in a substantial fraction of events. The X-ray lightcurves  show features like flares and a long-lasting plateau phase, whose origin is, at present, not clearly understood. This open question will be addressed in more detail in Section \ref{sec:X-ray-afterglow}.

The afterglow photons arises from the dissipation of the outflow kinetic energy. The supersonic flow of plasma interacts with the ambient developing a forward shock, where particle are accelerated by the Fermi acceleration mechanism \citep{Fermi1949} and magnetic field is amplified, likely due to the Weibel instability \citep{Weibel1959}. In such process, charged particles repeatedly cross the shock front gaining an energy proportional to the velocity of the shock front and forming a population of accelerated particle with a power-law distribution of Lorentz factor $N_{\gamma} \propto \gamma^{-p}$, where the index $p$ is the power-law index of the distribution, expected to assume a value $p\sim 2.5$. These charges than radiates via syncrotron process, generating the observed emission \citep{Sari1998}.

Along with the forward shock crossing the ambient medium, a reverse shock, propagating backward through the afterglow is expected to form if the jet, at this distance, is not strongly magnetized (\emph{i.e.} $\sigma \ll 1$) (\citealt{Meszaros1993,Sari1999}, see also \citealt{KuZh2015} for a review).

As long as the outflow is in relativistic motion, the radiation, emitted isotropically in the rest frame, is beamed within an angle $\theta_{\rm beam} \sim \Gamma^{-1}$ in the observer frame. While the jet decelerates and $\Gamma$ drops, the beaming angle increases and a wider portion of the emitting surface becomes visible to the observer. When $\theta_{\rm beam} \ge \theta_j$ the entire emission surface becomes visible to the observer, which means that there are no more de-beamed photons that can be revealed by a further increase in $\theta_{\rm beam}$ and the flux start to drop faster, as $\sim t^{-2}$, in the whole \ac{EM} spectrum \citep{Rhoads1999}. This achromatic feature, known as \emph{jet break}, has been one of the strongest evidence pointing towards the collimated nature of \ac{GRB} outflows, and, when observed, the time at which it occurs allows to measure the jet collimation angle $\theta_j$ \citep{SPH99}. This information, in turn, is fundamental to measure the true energetic of the $\gamma$-ray prompt emission \citep{Frail2001}.

\subsubsection{Open Problems}

In this section we address two of the principal open problems of \acp{GRB}: the physics behind the prompt emission spectra and the origin of the afterglow X-ray lightcurve.

\paragraph{Prompt Emission: The mystery of the GRB spectra}

The prompt emission spectrum indicates that the radiative processes is non-thermal. In particular, most of the observed \ac{GRB} spectra are modelled as two power laws smoothly connected at the peak energy in the flux spectrum (in units of $erg/cm^{2}/s$) \citep{Band1993,Preece1998,Frontera2000,Ghirlanda2002}. The power-law tail above the peak energy $\propto E^{-0.5}$, well resolved for the brightest \acp{GRB}, is prolonged in the energy range from hundreds of keV to sub-GeV range. It clearly supports the idea that the \ac{GRB} spectrum most probably originates from a power-law distributed population of charged particles (\emph{i.e.} $dN/d\gamma \propto \gamma^{-p}$, where $\gamma$ is their random Lorentz factor), e.g. electrons accelerated in shocks \citep{Rees1994,Piran1999}. 

The most straightforward process that could efficiently release the energy of charged electrons into the radiation is the synchrotron mechanism, which works both in the magnetic reconnection and internal shock dissipation scenarios. 
Therefore, the leptonic synchrotron radiation model to interpret the prompt emission fits the general paradigm of the internal dissipation of the jet. Its main advantage is to explain the non-thermal nature of the observed \ac{GRB} spectra by invoking a single and efficient radiative process. 

In a simple scenario, a population of accelerated electrons are 
injected in a single shot into the emitting region with a given averaged magnetic field $B$. This idealization is representative of the internal shocks scenario, where the collision between two shells produce shocks which propagate into each of the shell resulting in the charged particles acceleration and in the amplification of the pre-existed magnetic field. While the physical system seem to be oversimplified, the synchrotron model has a solid prediction of the spectral index of the prompt emission below its peak energy, independently from the exact shape of the accelerated particles. The reason resides in the spectrum of the single particle, which is characterized by a power-law segment, growing as $\propto E^{4/3}$, followed at higher energy by an exponential cut-off. When a power-law distribution of the emitters is considered, the low energy spectrum is dominated by the contribution of the particles with the lowest $\gamma$ resembling a single particle spectrum with the typical $E^{4/3}$ behavior. Above the peak instead, the spectrum is given by the envelope of the peaks of the single particle spectra with different $\gamma$, such that the shape of the spectrum is a power-law with a slope dependent on $p$. The lowest $\gamma$ of the particles distribution does not coincide in general with the lowest $\gamma$ of the initial distribution, which we call $\gamma_m$, because the charges loose energy by synchrotron emission at a rate $\propto \gamma^2$. After a certain time the emitters will populate states with $\gamma_c < \gamma < \gamma_m$ with a distribution like $dN/d\gamma \propto \gamma^2$, where $\gamma_c$ is a minimum Lorentz factor determined by the cooling, and the states above $\gamma_{m}$ with $dN/d\gamma \propto \gamma^{-p}$. The spectrum generated by such a distribution is a power-law $\propto E^{4/3}$ below $E_c$ (characteristic synchrotron energy for a particle with $\gamma = \gamma_c$), a power-law $\propto E^{1/2}$ between $E_c$ and $E_m$ (same of $E_c$ for $\gamma = \gamma_m$) and a power-law 
$\propto E^{1 -p/2}$ above $E_m$ \citep{RybickiLightman1986, Sari1998, Ghisellini2013}. 

The main assumption here was that the electrons' cooling time via synchrotron losses $t_{c}$ is much shorter than the dynamical time scale of the emitting region $t_{R}$ (i.e. $t_{c}<<t_{R}$), or alternatively, the integration time of the observed spectrum. This assumption is a requirement for an efficient radiation of the energy deposited in the electron population. It was shown, that the above-mentioned scenario, known as \emph{fast cooling regime} is the expected  regime in the  internal shocks model \citep{Ghisellini2000}. In this regime, the break at $E_m$ occurs in the hard X-rays, while the break at $E_c$ occurs at very low frequencies (\emph{e.g} radio). 
Thus, from X-ray to $\gamma$-ray the spectrum is expected to be a broken power-law with a single break at $E_m$.

The crisis of this scenario rises from the comparison of the low-energy photon indices with the expectation from the fast cooling regime of the synchrotron radiation. While the fast cooling regime predicts a spectrum of $\propto E^{1/2}$ below the peak energy, the typical observe spectra return $\propto E$  \citep{Tavani1996,Cohen1997,Crider1997,Preece1998}. Steeper spectra are also observed. Even more, there is some amount of \ac{GRB} spectra that are steeper than the steepest possible low-energy tail predicted in the synchrotron radiation model,  
namely $\propto E^{4/3}$ (see \emph{e.g.} \citealt{Acuner2019,Yu2019}).

This problem has been widely discussed in the literature. The proposed solutions
can be classified into two types: models which invoke emission mechanisms different than synchrotron radiation, and models which propose modifications to the basic synchrotron scenario. Among the first class of models, we recall scenarios invoking reprocessed emission, i.e. via Comptonization below the \ac{GRB} fireball reaches its transparency, mixed thermal plus non-thermal processes and inverse Compton reprocessing of softer photon field (\emph{e.g.} \citealt{Liang1997,Blinnikov1999,Rees2005,Giannios2006,Peer2008,Beloborodov2010,Chhotray2015,Vurm2016,Bhattacharya2020}). For the second class of models (studies that  consider synchrotron radiation above the photosphere) effects producing a hardening of the low-energy spectral index have been invoked, such as (1) effect of an energy dependent inverse Compton radiation (Klein-Nishina regime)
\citep{Derishev2001}, or self-absorption effect \citep{Lloyd2000}, (2) effects of the magnetic field profile within the emitting region \citep{Uhm2014}, (3) the nonuniform small-scale magnetic fields \citep{Medvedev2000}, (4) adiabatic cooling of electrons \citep{Geng2018,Panaitescu2019}, (5) re-acceleretation and slow or balanced heating of the cooling electrons \citep{Kumar2008,Asano2009,Xu2018,Beniamini2018} (7) supercritical regime of hadronic cascades, i.e. a rapid energy conversion from protons to pions and muons triggered by the $\gamma$-ray photons \citep{Petropoulou2014}.

All these proposals suggest specific configuration of the physical environments in which the observed emission is produced. Therefore, a clear identification of the radiative processes shaping the observed prompt emission spectra would be able of establishing the physics of the relativistic jets responsible for GRBs, i.e. the jet composition and its dissipation processes. In spite of all theoretical efforts, there is still no consensus on the origin of the prompt emission. 

In recent years, there were some additional observational and theoretical efforts aiming to resolve the dominant radiative processes responsible for the production of the GRB spectra. Among them, one intriguing discovery \citep{Oganesyan2017} was made in the broad-band studies of the prompt emission spectra. It turned out, that the prompt emission spectra extended down to soft X-rays, require a presence of an additional power-law segment. Once this break is included the spectral indices become consistent with the predicted synchrotron emission spectrum in the fast cooling regime. Namely, the spectrum below the break energy (few keV) is consistent with the synchrotron spectrum from a single electron $\propto E^{4/3}$, while above they are in agreement with the fast cooling segment $\propto E^{1/2}$. These findings were later confirmed with a different instrumentation and  with the break energy at larger energies ($\sim$ 100 keV) \citep{Ravasio2019}. The synchrotron radiation model in a marginally fast cooling regime \citep{Daigne2011,Kumar2008,Beniamini2013}, i.e. when electrons cool down very little ($\gamma_{c} \sim \gamma_{m}$) in a single observed time-window, was confirmed by the direct application of the synchrotron model to the \ac{GRB} data and by the early optical data, being consistent with $\propto E^{4/3}$ \citep{Oganesyan2019}. This scenario, suggested by the data, is quite challenging the single-shot acceleration model (where electrons are accelerated in a single episode) since it requires that electrons involved to produce the observed radiation, are somehow kept from efficient and fast cooling. Additionally, the parameter space of \ac{GRB} emitting side is non-trivial. It requires that the radiation is produced in weak magnetic fields ($B$ of few $G$), at relatively large radii ($R>10^{16}$ cm) which contradicts the observed variability time-scale of $t_{ang}\simeq R_{\gamma}/2c\Gamma^{2} \sim 0.1$ s for a reasonable range of bulk Lorentz factors $\Gamma \sim 100-300$. Several modifications to the standard model, i.e. to the single-shot acceleration of electrons emitting from a surface of a jet with an angular size of $\theta>>1/\Gamma$, could in principle provide this regime of radiation. Some models invoke continuous acceleration of electrons, and/or emission from large radii, invoking mini-jets to explain the short time variability \citep{Narayan2009}. 
However, mini-jets are likely to produce Comptonization\footnote{because a big fraction of the whole \ac{GRB} radiation is released in a region much smaller than the transverse size of the jet, so the radiation energy density is much higher than the case in which the entire surface emits \citep{Ghisellini2020}}, which is not observed in the spectra. 
Another possibility is that there is a continuous heating/re-acceleration of the electrons. If the acceleration time-scale becomes comparable with the electrons cooling time, then the synchrotron cooling frequency can be kept close to the peak energy, since particles are not allowed to cool down. This scenario was proposed in different dissipation models as a way to explain the observed spectral breaks in the X-rays \citep{Beniamini2018,Gill2020}. The re-acceleration of particles would require the presence of small scale turbulence and/or the closeby magnetic reconnection islands.

Recently \cite{Ghisellini2020} pointed out that is quite challenging to find a self-consistent and "comfortable" parameter space for models invoking electrons as the synchrotron emitters and proposed a model in which the synchrotron emission it is due to protons. In this \emph{proto-synchrotron} scenario the observed marginally fast-cooling spectrum can be explained even at low emitting ragion size (\emph{e.g. $R \sim 10^{13}\, \rm  cm$}) and this can in principle solve the tension with the short time variability requirement.

At present the puzzle of the prompt emission still remains unsolved. We do not have a complete understanding of the initial composition of the relativistic jets of GRBs, and the dissipative mechanisms operating in them. It is quite intriguing to unveil the clear picture of how the GRB jets are formed, 
accelerate and dissipate prior to their interaction with the interstellar medium. Therefore, GRBs open an interdisciplinary laboratory, where the emerged magneto-hydrodynamics, blast wave physics and and plasma physics could find its platform to face the astrophysical observables.    

\paragraph{Afterglow: The mystery of X-ray emission}
\label{sec:X-ray-afterglow}

Before the launch of {\it the Swift} satellite in 2004 \ac{GRB} afterglows were mainly observed at relatively late times. 
 The observed light curves were consistent with simple power-law in time, in agreement with the basic afterglow theory. However, the X-Ray Telescope (XRT) on board of {\it the Swift} shed new light onto the afterglow emission, revealing in the energy range of 0.3-10 keV complex behaviours which deviate from the simple blast wave deceleration profile \citep{Tagliaferri2005, Nousek2006,Zhang2006,OBrien2006}. 

The earliest phase of the \ac{GRB} X-ray emission consist in a steep decay. The flux temporal index $\alpha$ ($F \propto t^{-\alpha}$) is much larger than 2, which means that the emission fades much faster than predicted by the classical afterglow theory.
Moreover, after the steep decay, several X-ray afterglows show a plateau phase characterized by a shallow ($\alpha \sim 0.5$) temporal decay that can last for hours. This phase can be followed by an abrupt flux suppression or by a new decay phase, which resembles the standard afterglow decay. 
This broken power-law behavior observed in X-ray is inconsistent with the simple forward shock scenario. Afterglows presenting this feature are all but rare, since it is observed in $\sim\,80\,\%$ of \ac{LGRB} \citep{Evans2009, Margutti2013, Melandri2014} and $\sim\,50\,\%$ of $\ac{SGRB}$ \citep{Rowlinson2013, DAvanzo2014}. 

At late time ($\rm >10^{3}\, s$), X-ray flares are often observed \citep{Chincarini2010}. Their temporal behavior is similar to the prompt emission pulses. Therefore, if produced by internal energy dissipation, late-time X-ray flares require a re-activation of the \ac{GRB} central engine.

The X-ray afterglow without the above features behaves as the standard theory predict; a decay phases with $\alpha \sim 1$, followed by a late time fading with $\alpha \sim 2$ when the jet break occur.

The reason why the X-ray lightcurves behaves so differently with respect to the lightcurves in the other energy ranges, while standard afterglow theory would predict a similar trend, is still an open issue in the field.

At present, the most understood feature is the steep decay, which is interpreted as the tail of the prompt emission and, as such, not related to the forward shock that powers the afterglow. This occurs because the emission region (approximated with a surface) is curved, and the photons emitted at the same time in different parts of the region reach the observer at different times. 
Considering the observer is supposedly alligned with the jet axis, the photons emitted close to the axis arrive earlier than those at higher latitude. Moreover, since the jet is expanding relativistically, the emission is beamed around the (radial) direction of motion. This means that photons emitted farther from the axis will be more de-beamed from the observer line of sight with respect to those emitted close to the axis. Since the latters arrive before the formers, the observer detects a flux rapidly fading in time. This effect, usually referred as \emph{high latitude emission} (HLE) is able to account for the X-ray steep decay \citep{Fenimore1996,Kumar2000}. 

 On the contrary, the interpretation of X-ray plateaus is still debated. The more explored scenario invokes the presence of a long lasting central engine, that remains active for hours after the \ac{GRB} prompt emission \citep{Rees1998,Zhang2006}.
 Since such a long period of activity is difficult to be attained by an accreting \ac{BH} \citep[but see][]{KumarSci2008}, millisecond magnetars have been proposed as a more suitable central engine candidate \citep{Dai1998,Zhang2001}. As already mentioned in Section \ref{sec:GRB_centralengine}, a millisecond magnetar loose rotational energy via magnetic dipole spindown emission. This energy can be injected into the forward shock (external dissipation), or dissipated in proximity of the central engine (internal dissipation), sustaining in this way the X-ray emission. 
 
 A key feature of the magnetar central engine is that the magnetic dipole spindown emission is characterized by a constant luminosity up to a certain time, the characteristic spindown time when half of the energy is radiated, and then the luminosity is expected to decline as $t^{-2}$. Interestingly, for a millisecond magnetar, the spindown timescale is of the same order of magnitude of the typical plateau duration. Therefore, this well defined temporal behavior can straightforwardly account for the plateau in X-ray afterglow lightcurves and the successively decay phase. Moreover, if the magnetar is metastable, it may collapse into a \ac{BH} when the spindown deprives it from centrifugal support. When this happens, the energy outflow ceases abruptly, and if the X-ray radiation is generated by internal dissipation this can explain the sharp drop in flux sometimes observed (see \emph{e.g.} \citealt{Troja2007,Sarin2020}). The magnetar model can successfully fit those \ac{GRB} lightcurves that present a plateau \citep{Lyons2010, DallOsso2011, Bernardini2012, Bernardini2013, Rowlinson2013} and can also account for an observed anticorrelation between plateau luminosity and duration \citep{Dainotti2008}. 
 
 Although promising, the magnetar central engine, as already explained, cannot guarantee the formation of the jet, due to the large amount of baryons in the surrounding medium expected when a long-lived \ac{NS} results from the merger.
 Since the jet is essential to explain both the prompt emission and the afterglow, this constitute a very important issue of the magnetar model. 
 
 A different class of models that avoid the problems of the long lasting central engines invoke a more refined description of the the \ac{GRB} jet. \citet{BeniaminiDuque2020} recently developed a model in which the X-ray plateau is generated by the forward shock driven by a structured jet\footnote{A structured jet is a jet where the velocity and energy of the fluid have a non-constant profile with respect to the angular distance measured from the jet axis.}, when the observer is located slightly outside the jet axis. Another model have been proposed recently by \citet{Oganesyan2020}, in which both the steep-decay and the plateau in X-ray can be explained by the prompt high-latitude emission when the jet is structured. Therefore, according to \citet{Oganesyan2020}, the X-ray plateau is part of the tail of the prompt emission, while the forward shock is responsible of the optical and radio afterglow. This scenario is able to explain why the afterglow follows the standard theory in the optical band and  often it does not in the X-rays.
 
Concerning the X-ray flares, a late time activity of the central engine is often invoked. This may be provided either by a magnetar central engine \citep{Dai2006, Metzger2011, Bernardini2013} either by mechanisms similar to those responsible for the prompt emission or by dissipation of magnetic field as occurs in Galactic magnetars or by a \ac{BH} accreting at late time \citep{PeArZh2006, PrZh2006, LeeRamirezRuiz2009, DallOsso2017}.

\subsection{Kilonova}
\label{sec:kilonova}

As it was already stated, during \bns or a \nsbh coalescence an amount of NS matter is expected to be unbound from the central remnant and ejected with mildly relativistic velocity. Since these expanding ejecta, as formed by \ac{NS} decompressed matter, are neutron rich, they constitute a natural environment for r-process nucleosynthesis. This chain of nuclear reactions is characterized by the rapid capture of a number of free neutrons by neutron rich nuclei, leading to the formation of very heavy elements. Due to the excess of neutrons, these elements are metastable and they eventually decay (mainly through $\beta$-decay) turning some neutrons into protons.\footnote{Here the radioactive decay is slower than the rate of neutron capture. When the decay is instead faster, the nucleosynthesis occurs via the \emph{s-process} (or ``slow'' process)}. 
The radioactive decay of the newly synthesized heavy nuclei deposits energy within the ejecta via non-thermal particles such as $\gamma$-rays, $\alpha$ and $\beta$ particles, and fission fragments, which then thermalizes efficiently and turns into ejecta internal energy. This energy excess is finally radiated away and produces the thermal transient which is commonly referred to as \emph{kilonova} or \emph{macronova} \citep{LP98, Kulkarni:2005jw, Metzger10, Metzger2017} \footnote{Some aspects of mass ejection and the possibility of an \ac{EM} emission powered by the radioactive nuclides had been also anticipated in \citet{Blinnikov1990}}. 

Within the context of a single component isotropic model (a model characterized by a single isotropic ejecta with uniform composition) it is possible to obtain an estimate for the main \ac{KN} observational features, such as the peak luminosity, time at which the peak occur and brightness temperature at the peak. In this context the governing parameters are three: the mass $M_{\rm ej}$, the velocity $v_{\rm ej}$ and the opacity $k$ of the the ejecta. 
A simple realistic model can be built using these three parameters. In fact, the first parameter, representing the total amount of matter undergoing radioactive decay,  constitutes the "fuel" of the \ac{KN}, while all the three parameters determine the characteristic photon diffusion timescale.
Taking into account an homogeneous composition and assuming that the peak of the emission occurs when the timescale of expansion equals the diffusion timescale \citep{Arnett1982}, it is possible to obtain the following equations \citep{Metzger10}:

\begin{align}
t_{\rm peak}&=\Bigl(\frac{3kM_{\rm ej}}{4\pi c v_{\rm ej}}\Bigr)^{1/2}\simeq2.7\,\rm days\,\Bigl(\frac{M_{\rm ej}}{10^{-2}M_\odot}\Bigr)^{1/2}\Bigl(\frac{v_{\rm ej}}{0.1c}\Bigr)^{-1/2}\Bigl(\frac{k}{\rm cm^2g^{-1}}\Bigr)^{1/2}\notag\\
L_{\rm peak}&=M_{\rm ej}\dot{\epsilon}_{n}(t_{\rm peak})\simeq5\times10^{40}\,\frac{\rm erg}{\rm s}\epsilon_{\rm th}\Bigl(\frac{M_{\rm ej}}{10^{-2}M_\odot}\Bigr)^{0.35}\Bigl(\frac{v_{\rm ej}}{0.1c}\Bigr)^{0.65}\Bigl(\frac{k}{\rm cm^2g^{-1}}
\Bigr)^{-0.65}\label{eq:KNtoymodel} \\
T_{\rm peak}&=\Bigl(\frac{L_{\rm peak}}{4\pi\sigma R^{2}_{\rm peak}}\Bigr)^{1/4}\simeq 3460\,\rm K\,\epsilon^{1/4}_{\rm th}\Bigl(\frac{M_{\rm ej}}{10^{-2}M_\odot}\Bigr)^{-0.17}\Bigl(\frac{v_{\rm ej}}{0.1c}\Bigr)^{-0.09}
\Bigl(\frac{k}{\rm cm^2g^{-1}}\Bigr)^{-0.41}\notag
\end{align}
where $\dot{\epsilon}_n$ is the specific radioactive heating rate, $\epsilon_{\rm th}\le1$ is the thermalization efficiency and $R_p=v_{\rm ej}t_p$ is the ejecta typical size at time $t_p$ . 

The opacity is determined by the composition of the matter. It depends on whether elements of the group of lanthanides are synthesized or not during r-process nucleosynthesis. These elements, due to the numerous transition lines in the optical bands, give rise to an opacity which can be even two order of magnitude higher than that of the iron group elements. If they are present, they dominate the net opacity of the ejecta \citep{BaKa2013, TaHo2013}. 
Their synthesis depends on the electron fraction of the ejected matter, which is defined as $Y_e \equiv n_p/(n_n + n_p)$ - where $n_p$ and $n_n$ are the proton and neutron number densities - and measures how much the matter is neutron rich. In particular, considering a threshold of $Y_e \simeq 0.25$, low $Y_e$ ensures the synthesis of lanthanides, while high $Y_e$ prevents it \citep{Wu2016}. Equations \ref{eq:KNtoymodel} show that the effect of increasing opacity is to shift the peak of the lightcurve at later time, to make the emission fainter, and to decrease its peak temperature making the transient redder in color. 

\begin{figure}
	\includegraphics[width = 0.5\columnwidth]{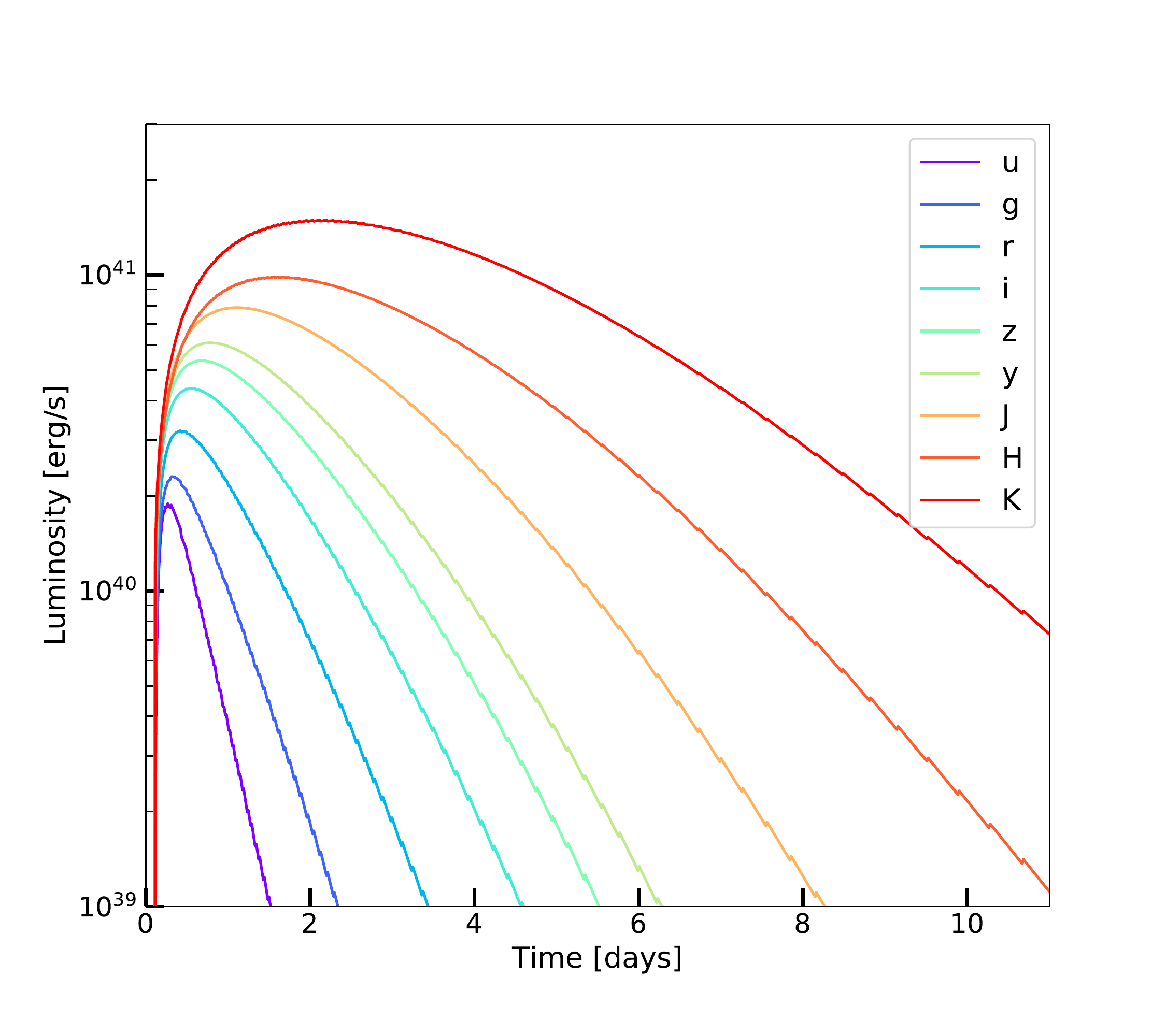}
	\includegraphics[width = 0.47\columnwidth]{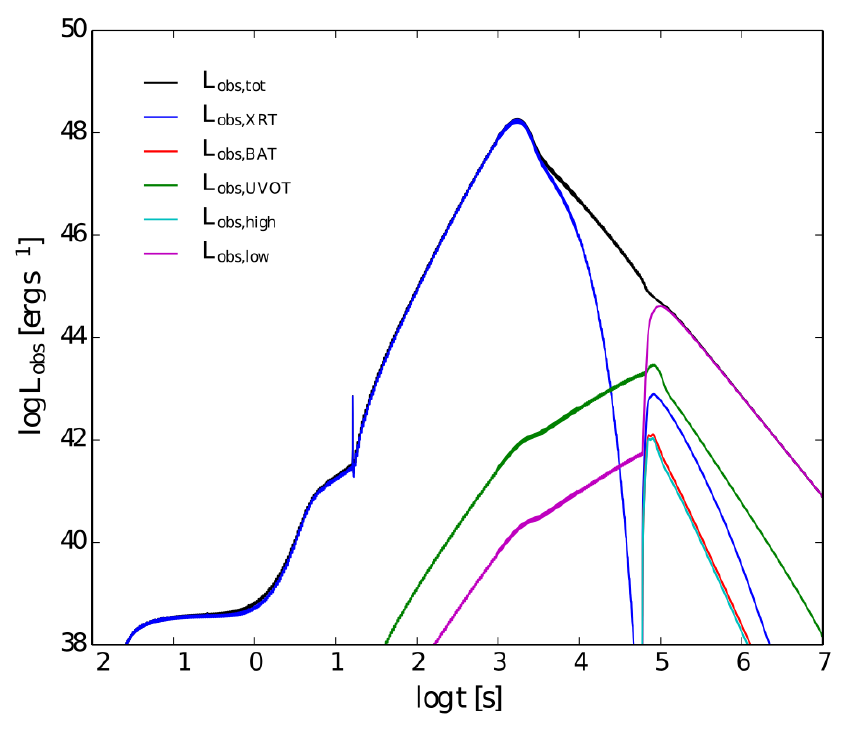}
	\caption{\emph{Left}: KN lightcurve in different photometric filters (optical and \ac{NIR}) from a single component model characterized by $M_{\rm ej} = 0.01\,M_\odot$, $v_{\rm ej} = 0.1\,c$ and $k = 10\,\rm cm^2/g $. Lightcurves obtained with the code in \url{https://github.com/mcoughlin/gwemlightcurves}. \emph{Right}: \ac{SDPT} lightcurve from the fiducial model in \citet{SiegelCiolfi2016b} characterized by $M_{\rm ej} = 5\times 10^{-3}\,M_\odot$ and $B= 10^{16}\,\rm G$. In this case the \ac{NS} does not collapse. Black curve represents the bolometric luminosity, the dark blue curve is the lightcurve in the energy range of \emph{Swift}-XRT (0.3-10 keV), the red curve in \emph{Swift}-BAT range (15-150 keV), the green curve in \emph{Swift}-UVOT range (170-650 nm). Light blue and purple curves represent lightcurves above BAT range and below UVOT range, respectivelly. Figure adapted from \citet{SiegelCiolfi2016b}.}
	\label{fig:kilonova_model}
\end{figure}

In the left panel of Fig. \ref{fig:kilonova_model} it is shown an example 
of a simple kilonova model outlined in \citet{Metzger2017} characterized by mass of the ejecta $M_{\rm ej} = 0.01\,M_\odot$, $v_{\rm ej} = 0.1\,c$ and $k = 10\,\rm cm^2/g$.

Although the single component model is instructive to understand the qualitative impact of the parameters on the \ac{KN} lightcurve and to have an indication of the brightness and spectral range of the emission (at an order-of-magnitude level), it is too simple for a comparison with the data. In fact, as it was anticipated in Sec. \ref{sec:coalescence}, hydrodynamical simulations showed that in a \ac{CBM} several ejection mechanisms occurr at different times leading to multiple ejecta components, with different mass, velocity and chemical composition (\emph{i.e.} opacity). This prediction was confirmed by the detection of AT2017gfo, whose lightcurve is very hard to explain with a single component model \citep{PianDavanzo2017, Tanaka2017, PeregoRadice2017}. 

In a \bns merger, an amount of matter of $M_{ej} \sim 10^{-3}-10^{-2}\,M_\odot$ is expelled via dynamical mass ejection during the merger itself with a velocity of $v_{\rm ej} \sim 0.1-0.3\, c$ and an electron fraction $Y_e \sim 0.05 -0.4$. Tidal ejecta are responsible for the lower $Y_e$ component and are located on the binary equatorial plane. Shock-driven ejecta, which cover the polar region, have higher $Y_e$, due to the fact that the matter is heated by the shocks and this leads to pair production and consequently to the capture of positrons by neutrons \citep{R15}.
In the case of \nsbh merger only the tidal component is present and the unbound mass can reach $M_{\rm ej} \sim 10^{-1}\, M_\odot$ \citep{R05, Foucart2014}. 

The presence of a (meta)stabe massive NS, only for \bns mergers, would produce further mass outflow in the form of neutrino-driven (e.g., \citealt{DesOttBurRos2009,PeRo2014,MaPe2015}) and/or magnetically-driven (e.g., \citealt{SieCioRez2014,CiolfiKastaun2017,CiolfiKalinani2020}) baryon loaded-winds, with a potentially high electron fraction due to the effect of neutrino irradiation from the NS itself, a mass of up to a few $\sim 10^{-2}\,M_\odot$, and a velocity of $\sim 0.1-0.2\, c$ (where the higher end can be achieved in the polar region in presence of a strong magnetic field; \citealt{CiolfiKalinani2020}). 

Furthermore, after the collapse to a BH for \bns systems or simply after merger for \nsbh systems, up to $40\%$ of the accretion disk mass can be ejected via MHD turbulence and neutrino heating within the disk itself (e.g., \citealt{Siegel2018,Fernandez2019}). These $\emph{disk wind ejecta}$ are characterized by a wide range of electron fractions ($Y_e\sim 0.1-0.5$) and low velocity $v_{\rm ej} \lesssim 0.1\, c$ (e.g., \citealt{PeRo2014, MaPe2015,FernandezQuataert2015, Wu2016, SiegelMetzger2017,Siegel2018,Fujibayashi2018, RadicePerego2018}). 

The presence of different matter components, with different mass, velocity and electron fraction, results in more complicated lightcurves with fast evolving bluer components (\emph{blue \ac{KN}}) and slower evolving redder components (\emph{red \ac{KN}}). Moreover, since each of the components have a different geometrical/angular distribution, the resulting \ac{KN} lightcurve also depend on the orientation of the system with respect to the observer (e.g., \citealt{ PeregoRadice2017, WoKo2018}).

\subsection{Spindown-powered transients}
\label{sec:SDPT}

When the outcome of a \bns merger is another \ac{NS} (stable or metastable), the newborn \ac{NS} is expected to spin very fast- with rotational periods of $\sim 1\,\rm ms$ and to have a strong magnetic field - up to $\sim 10^{16}\, G$ in the interior - due to amplification mechanisms. 
During the first $O(100)\, \rm ms$ the \ac{NS} is hot and in differential rotation. In this phase, the strong neutrino irradiation and the gradient of magnetic pressure generated by the field winding due the action of differential rotation are responsible for a baryon-loaded wind (see also previous Section). 
If the \ac{NS} is a stable \ac{NS} or \ac{SMNS} it does not collapse when the core differential rotation is quenched, and it has a sufficient amount of time to lose its huge amount of rotational energy ($O(10^{52}\,\rm erg)$) through magnetic dipole spindown emission \citep{Pacini1967}. The energy emitted at first as a Poynting flux is converted close to the source (by dissipation mechanisms not fully understood) in a electron-positron pair rich wind, analogous to a \emph{pulsar wind}, expanding with relativistic velocities\footnote{While this process has never been observed in a \bns merger scenario, this is what happens in the (less energetic) case of a \ac{PWN}, which is a photon-pair plasma nebula powered by the magnetic spin-down emission from a \ac{NS} formed after a stellar core-collapse. See \citet{PWN2006} for a review on the topic.}. 
When the fast pulsar wind encounters the previously emitted and much slower baryon wind it drives a shock through it increasing its temperature and boosting its expansion. The thermal radiation from these hot expanding ejecta, gives origin to a Spin-Down Powered Transient (\ac{SDPT}) which is expected to be observable in X-ray, UV and optical bands (e.g., \citealt{YuZhangGao2013,MePi2014,SiegelCiolfi2016a,SiegelCiolfi2016b}). 

So far, a compelling detection of this type of transients is still missing, although potential associations have been discussed in the literature (e.g., \citealt{Ciolfi2016,XueZheng2019}).
\acp{SDPT} are of particular interest since they are the only \ac{EM} counterparts of \ac{CBM} whose detection is a clear signature of the formation of a long-lived \ac{NS} after the merger. The formation of such a type of \ac{NS} would allow us to further constrain the \ac{NS} \ac{EOS}. 
Moreover, a SDPT detection would represent a secure way to distinguish a \bns merger from a \nsbh merger.

So far \ac{SDPT} have been studied only through 1D  semi-analytical approaches. These models consider a central \ac{NS}, radiating via magnetic dipole emission, embedded in a \ac{PWN} surrounded by an 
expanding spherical baryon wind. One of the first model of this kind  was developed by \citet{YuZhangGao2013}, who found a peak luminosity in the range $10^{44}-10^{45}\,\rm erg/s$ occurring $10^4-10^5\, \rm s$ after the merger. Later \citet{GaoDing2015} applied this model to the \ac{SGRB} GRB080503 to explain feature observed in its lightcurve . A similar model was developed by \citet{MePi2014}, who provided a deeper focus on \ac{PWN} physics in particular adding a self-consistent treatment of \ac{PWN} opacity and baryon ejecta degree of ionization and albedo.
They predict in this way a dimmer transient with respect to the \citet{YuZhangGao2013} model with peak luminosity in range $10^{43}-10^{44}\,\rm erg/s$ peaking at $10^4-10^5\, \rm s$ after merger. 
\citet{SiegelCiolfi2016a} advanced further the complexity of the treatment by including a relativistic dynamics, and starting the evolution from the baryon wind ejection, thus before the pulsar wind launching. In this way they followed three phases of the transient evolution: the baryon wind emitting phase, a second phase  characterized by the pulsar wind and baryon wind interaction and the crossing of the shock through the latter, and a final phase in which the baryon wind expands under the pressure of an inner \ac{PWN}. 
They predict that the transient peaks in the soft X-rays at $\sim 10^2-10^4\,\rm s$ after the merger, with a luminosity ranging in the interval $10^{46}-10^{48}\rm erg/s$ \citep{SiegelCiolfi2016b}.
In the right panel of Fig. \ref{fig:kilonova_model} an example of SPDT model from \citet{SiegelCiolfi2016b} is shown.

\section{Final Remarks}
\label{sec:final}

The merger of a binary system composed by two \acp{NS} or a \ac{NS} and a \ac{BH} is a powerful source of gravitational radiation that can also be followed by an intense emission of photons across the whole \ac{EM} spectrum. 
So far, only one event of this kind, the \bns merger named GW170817, has been observed through these two different messengers. However, the information we gained by this single detection increased considerably our understanding in many different fields of astrophysics and fundamental physics. 

The observation of a flash in $\gamma$-ray \citep{Goldstein2017, Savchenko2017, GW170817MMA} along with a collimated outflow moving relativistically from the centre of explosion \citep{MooleyDeller2018, Ghirlanda2019} allowed us to confirm that (at least some) \acp{SGRB} are generated after a \bns merger.
The multiwavelenght observation of the afterglow showed that \ac{GRB} relativistic jet are structured \citep{TrojaPiro2017, TrojaPiro2018, AlexanderMargutti2018, DavanzoCampana2018}. 

The first \bns observation coincides also with the observation of a \ac{KN}. 
The brightness and properties of the \ac{KN} demonstrated that \acp{CBM} are favorable site of r-processes, which is the synthesis channel with whom the heaviest elements form in the universe.

GW170817 has been also used to measure the Hubble constant \citep{Fishbach2019}, and the limits on the tidal deformability of the \acp{NS} have been used to constrain the dense matter \ac{EOS} \citep{LVC2018_tidal}. 

Although the scientific outcome of GW170817 have been wide, many open questions concerning \acp{CBM} and their \ac{EM} counterparts still remains unsolved. For example, at present, it is not clear how much the magnetic field can be amplified during a \bns merger (see, e.g., \citealt{Ciolfi2020b} for a recent review). The magnitude of the magnetic field, in turn, impacts the amount of matter ejected during the merger, the possibility to launch a jet and/or to form a magnetar as the result of the coalescence (and thus whether \acp{SDPT} occurs in nature or not). 

Concerning \acp{GRB}, many aspects of their physics are still not clear: is a magnetar able to launch a \ac{GRB} jet? Is the jet more similar to a hot fireball or to a cold Poynting flux dominated outflow? What are the dissipation mechanisms responsible for converting the jet kinetic energy in $\gamma$-ray radiation? What is the origin of the X-ray plateaus and X-ray flares often observed to occur in \ac{GRB} afterglows? These are among the principal questions that astrophysicist need to address in the near future. Plasma physics is expected  to play a key role in addressing them. 

In the near future, the gravitational interferometers, the X-ray surveys such as eROSITA \citep{eROSITA2012}, the wide field X-ray telescopes such as THESUS-SXI \citep{Theseus2018,Stratta2018}, and the optical surveys such as the VEra Rubin Observatory \citep{LSST2019}, are going to provide us more and more joint \ac{GW}-\ac{EM} detections of \acp{CBM}. This represents a unique opportunity to unveil the physical mechanism at the base of the most energetic events in the Universe and to study extreme environments characterized by intense gravity, strong magnetic fields, high densities and temperatures which cannot be reproduced in terrestrial laboratory. 
\acp{CBM} observations will allow us to get new insights into the physics of astrophysical plasmas.

\begin{acknowledgements}
The authors thank the PLUS collaboration (Coordinator: Giovanni Montani). We thank Alin Panaitescu for the plot in Fig. \ref{fig:afterglow_LC}. We thank Gabriele Ghisellini and Om Sharan Salfia for fruitful discussion. SA acknowledges the PRIN-INAF "Towards the SKA and CTA era: discovery, localization and physics of transient sources" and the ERC Consolidator Grant
“MAGNESIA” (nr.817661).  SA acknowledges the GRAvitational Wave Inaf TeAm -
GRAWITA (P.I. E. Brocato). GO and MB acknowledge financial contribution from the
agreement ASI-INAF n.2017-14-H.0. MB acknowledge financial support from MIUR (PRIN 2017 grant 20179ZF5KS). 
\end{acknowledgements}

\bibliographystyle{jpp}

\bibliography{references}

\end{document}